\pdfoutput=1
\documentclass[12pt,a4paper]{article}

\usepackage{ifthen} 
\newboolean{pdflatex}
\setboolean{pdflatex}{true} 

\newboolean{articletitles}
\setboolean{articletitles}{true} 

\newboolean{uprightparticles}
\setboolean{uprightparticles}{false} 

\newboolean{inbibliography}
\setboolean{inbibliography}{false} 

\def\paperauthors{LHCb collaboration}
\def\paperasciititle{Helium identification with LHCb} 
\def\papertitle{Helium identification with LHCb} 
\def\paperkeywords{{High Energy Physics}, {LHCb}} 
\def\papercopyright{\the\year\ CERN for the benefit of the LHCb collaboration} 
\def\paperlicence{CC BY 4.0 licence}
\def\paperlicenceurl{https://creativecommons.org/licenses/by/4.0/}

\usepackage[top=1in, bottom=1.25in, left=1in, right=1in]{geometry}

\usepackage{microtype}
\usepackage{multirow}
\usepackage{lineno}  
\usepackage{xspace} 
\usepackage{caption} 

\usepackage{graphicx}  
\usepackage{color}
\usepackage{colortbl}
\graphicspath{{./figs/}} 
\usepackage{overpic} 

\usepackage{amsmath} 
\usepackage{amssymb}
\usepackage{amsfonts}
\usepackage{pifont} 
\usepackage{upgreek} 

\newcommand*\patchAmsMathEnvironmentForLineno[1]{%
\expandafter\let\csname old#1\expandafter\endcsname\csname #1\endcsname
\expandafter\let\csname oldend#1\expandafter\endcsname\csname
end#1\endcsname
 \renewenvironment{#1}%
   {\linenomath\csname old#1\endcsname}%
   {\csname oldend#1\endcsname\endlinenomath}%
}
\newcommand*\patchBothAmsMathEnvironmentsForLineno[1]{%
  \patchAmsMathEnvironmentForLineno{#1}%
  \patchAmsMathEnvironmentForLineno{#1*}%
}
\AtBeginDocument{%
\patchBothAmsMathEnvironmentsForLineno{equation}%
\patchBothAmsMathEnvironmentsForLineno{align}%
\patchBothAmsMathEnvironmentsForLineno{flalign}%
\patchBothAmsMathEnvironmentsForLineno{alignat}%
\patchBothAmsMathEnvironmentsForLineno{gather}%
\patchBothAmsMathEnvironmentsForLineno{multline}%
\patchBothAmsMathEnvironmentsForLineno{eqnarray}%
}

\usepackage{hyperxmp}

\usepackage[pdftex,
            pdfauthor={\paperauthors},
            pdftitle={\paperasciititle},
            pdfkeywords={\paperkeywords},
            pdfcopyright={Copyright (C) \papercopyright},
            pdflicenseurl={\paperlicenceurl}]{hyperref}

\usepackage[colorinlistoftodos,textsize=scriptsize]{todonotes}

\usepackage[bottom,flushmargin,hang,multiple]{footmisc}

\usepackage[all]{hypcap} 

\usepackage{cleveref} 
\Crefname{figure}{Fig.}{Figs.}
\Crefname{tabular}{Tab.}{Tabs.}
\Crefname{section}{Sect.}{Sects.}
\Crefname{reference}{Ref.}{Refs.}

\usepackage{xspace} 
\usepackage{upgreek}

\def\lhcb   {\mbox{LHCb}\xspace}

\def\alice  {\mbox{ALICE}\xspace}

\def\lhc    {\mbox{LHC}\xspace}

\def\velo   {VELO\xspace}
\def\rich   {RICH\xspace}

\def\ttracker {TT\xspace}
\def\intr   {IT\xspace}

\def\ot     {OT\xspace}

\def\MagUp {\mbox{\em Mag\kern -0.05em Up}\xspace}

\ifthenelse{\boolean{uprightparticles}}%
{

 \def\Ppi         {\ensuremath{\uppi}\xspace}

 \def\PDelta      {\ensuremath{\Delta}\xspace}                 
 \def\PXi         {\ensuremath{\Xi}\xspace}                 
 \def\PLambda     {\ensuremath{\Lambda}\xspace}                 
 \def\PSigma      {\ensuremath{\Sigma}\xspace}                 
 \def\POmega      {\ensuremath{\Omega}\xspace}                 
 \def\PUpsilon    {\ensuremath{\Upsilon}\xspace}
 \let\oldPi\Pi
 \def\PPi         {\ensuremath{\oldPi}\xspace}

 \def\PB      {\ensuremath{\mathrm{B}}\xspace}                 
                  
 \def\PD      {\ensuremath{\mathrm{D}}\xspace}

 \def\PK      {\ensuremath{\mathrm{K}}\xspace}

 \def\Pb      {\ensuremath{\mathrm{b}}\xspace}

 \def\Pe      {\ensuremath{\mathrm{e}}\xspace}

 \def\Pi      {\ensuremath{\mathrm{i}}\xspace}

 \def\Pp      {\ensuremath{\mathrm{p}}\xspace}

 \def\Ps      {\ensuremath{\mathrm{s}}\xspace}

 \def\thebaroffset{0.0em}
}
{

 \def\Ppi         {\ensuremath{\pi}\xspace}

 \mathchardef\PDelta="7101
 \mathchardef\PXi="7104
 \mathchardef\PLambda="7103
 \mathchardef\PSigma="7106
 \mathchardef\POmega="710A
 \mathchardef\PUpsilon="7107
 \mathchardef\PPi="7105
                  
 \def\PB      {\ensuremath{B}\xspace}                 
                  
 \def\PD      {\ensuremath{D}\xspace}

 \def\PK      {\ensuremath{K}\xspace}

 \def\Pb      {\ensuremath{b}\xspace}

 \def\Pe      {\ensuremath{e}\xspace}

 \def\Pi      {\ensuremath{i}\xspace}

 \def\Pp      {\ensuremath{p}\xspace}

 \def\Ps      {\ensuremath{s}\xspace}

 \def\thebaroffset{0.18em}
}
\newcommand{\offsetoverline}[2][\thebaroffset]{\kern #1\overline{\kern -#1 #2}}%

\makeatletter
\ifcase \@ptsize \relax
  \newcommand{\miniscule}{\@setfontsize\miniscule{4}{5}}
\or
  \newcommand{\miniscule}{\@setfontsize\miniscule{5}{6}}
\or
  \newcommand{\miniscule}{\@setfontsize\miniscule{5}{6}}
\fi
\makeatother

\DeclareRobustCommand{\optbar}[1]{\shortstack{{\miniscule (\rule[.5ex]{1.25em}{.18mm})}
  \\ [-.7ex] $#1$}}

\def\en         {{\ensuremath{\Pe^-}}\xspace}   
\def\ep         {{\ensuremath{\Pe^+}}\xspace}

\def\squark    {{\ensuremath{\Ps}}\xspace}

\def\bquark    {{\ensuremath{\Pb}}\xspace}

\def\pion   {{\ensuremath{\Ppi}}\xspace}

\def\pip    {{\ensuremath{\pion^+}}\xspace}
\def\pim    {{\ensuremath{\pion^-}}\xspace}

\def\kaon    {{\ensuremath{\PK}}\xspace}

\def\KorKbar {\kern \thebaroffset\optbar{\kern -\thebaroffset \PK}{}\xspace}

\def\Kp      {{\ensuremath{\kaon^+}}\xspace}
\def\Km      {{\ensuremath{\kaon^-}}\xspace}

\def\Kstarz  {{\ensuremath{\kaon^{*0}}}\xspace}

\def\D       {{\ensuremath{\PD}}\xspace}

\def\DorDbar {\kern \thebaroffset\optbar{\kern -\thebaroffset \PD}\xspace}
\def\Dz      {{\ensuremath{\D^0}}\xspace}

\def\Dp      {{\ensuremath{\D^+}}\xspace}
\def\Dm      {{\ensuremath{\D^-}}\xspace}

\def\DpDm    {\ensuremath{\Dp {\kern -0.16em \Dm}}\xspace}

\def\Dstarp  {{\ensuremath{\D^{*+}}}\xspace}

\def\B       {{\ensuremath{\PB}}\xspace}

\def\BorBbar {\kern \thebaroffset\optbar{\kern -\thebaroffset \PB}\xspace}

\def\Bd      {{\ensuremath{\B^0}}\xspace}

\def\BdorBdbar {\kern \thebaroffset\optbar{\kern -\thebaroffset \Bd}\xspace}

\def\Bs      {{\ensuremath{\B^0_\squark}}\xspace}

\def\BsorBsbar {\kern \thebaroffset\optbar{\kern -\thebaroffset \Bs}\xspace}

\def\Y#1S{\ensuremath{\PUpsilon{(#1S)}}\xspace}

\def\proton      {{\ensuremath{\Pp}}\xspace}

\def\Lz          {{\ensuremath{\PLambda}}\xspace}

\def\LorLbar     {\kern \thebaroffset\optbar{\kern -\thebaroffset \PLambda}\xspace}

\def\Lb           {{\ensuremath{\Lz^0_\bquark}}\xspace}

\def\to                 {\ensuremath{\rightarrow}\xspace}

\def\AT#1     {\ensuremath{A_{\mathrm{T}}^{#1}}\xspace}           

\def\C#1      {\ensuremath{\mathcal{C}_{#1}}\xspace}                       
\def\Cp#1     {\ensuremath{\mathcal{C}_{#1}^{'}}\xspace}                    
\def\Ceff#1   {\ensuremath{\mathcal{C}_{#1}^{\mathrm{(eff)}}}\xspace}        
\def\Cpeff#1  {\ensuremath{\mathcal{C}_{#1}^{'\mathrm{(eff)}}}\xspace}       
\def\Ope#1    {\ensuremath{\mathcal{O}_{#1}}\xspace}                       
\def\Opep#1   {\ensuremath{\mathcal{O}_{#1}^{'}}\xspace}                    

\newcommand{\nospaceunit}[1]{\ensuremath{\text{#1}}}       
\newcommand{\aunit}[1]{\ensuremath{\text{\,#1}}}       

\newcommand{\tev}{\aunit{Te\kern -0.1em V}\xspace}
\newcommand{\gev}{\aunit{Ge\kern -0.1em V}\xspace}
\newcommand{\mev}{\aunit{Me\kern -0.1em V}\xspace}
\newcommand{\kev}{\aunit{ke\kern -0.1em V}\xspace}
\newcommand{\ev}{\aunit{e\kern -0.1em V}\xspace}
 
\newcommand{\mevc}{\ensuremath{\aunit{Me\kern -0.1em V\!/}c}\xspace}
\newcommand{\gevc}{\ensuremath{\aunit{Ge\kern -0.1em V\!/}c}\xspace}
\newcommand{\mevcc}{\ensuremath{\aunit{Me\kern -0.1em V\!/}c^2}\xspace}
\newcommand{\gevcc}{\ensuremath{\aunit{Ge\kern -0.1em V\!/}c^2}\xspace}

\def\mm   {\aunit{mm}\xspace}

\def\mum  {\ensuremath{\,\upmu\nospaceunit{m}}\xspace}

\def\fb   {\ensuremath{\aunit{fb}}\xspace}
\def\invfb   {\ensuremath{\fb^{-1}}\xspace}

\def\ns   {\ensuremath{\aunit{ns}}\xspace}

\newcommand{\chisq}{\ensuremath{\chi^2}\xspace}

\newcommand{\chisqip}{\ensuremath{\chi^2_{\text{IP}}}\xspace}

\def\gsim{{~\raise.15em\hbox{$>$}\kern-.85em
          \lower.35em\hbox{$\sim$}~}\xspace}
\def\lsim{{~\raise.15em\hbox{$<$}\kern-.85em
          \lower.35em\hbox{$\sim$}~}\xspace}

\def\sqs   {\ensuremath{\protect\sqrt{s}}\xspace}

\def\pt         {\ensuremath{p_{\mathrm{T}}}\xspace}

\def\ptot       {\ensuremath{p}\xspace}

\newcommand{\dedx}{\ensuremath{\mathrm{d}\hspace{-0.1em}E/\mathrm{d}x}\xspace}

\def\geant      {\mbox{\textsc{Geant4}}\xspace}

\def\tell1  {TELL1\xspace}
\def\ukl1   {UKL1\xspace}

\def\adc    {ADC\xspace}

\newcommand{\ie}{\mbox{\itshape i.e.}\xspace}

\newcommand{\lhcborcid}[1]{\href{https://orcid.org/#1}{\hspace*{0.1em}\raisebox{-0.45ex}{\includegraphics[width=1em]{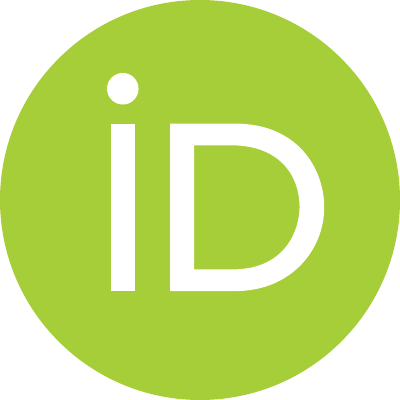}}}}

\usepackage[version=4]{mhchem} 

\def\He           {{\ensuremath{^3{\rm He}}}\xspace}

\def\Hefour       {{\ensuremath{^4{\rm He}}}\xspace}

\newcommand{\LLD}[1]{\mbox{\ensuremath{\Lambda_{\rm LD}^{#1}}}\xspace}
\newcommand{\LLDvelo}{\LLD{\rm \velo}}
\newcommand{\LLDveloR}{\LLD{{\rm \velo}\,R}}
\newcommand{\LLDvelophi}{\LLD{{\rm \velo}\,\phi}}
\newcommand{\LLDTT}{\LLD{\rm \ttracker}}
\newcommand{\LLDIT}{\LLD{\rm \intr}}

\newcommand{\LL}[1]{\mbox{\ensuremath{\mathcal L^{#1}}}\xspace}

\newcommand{\pp}{{\ensuremath{\proton\proton}}\xspace}

\newcommand{\PbPb}{{\ensuremath{\ce{Pb}\ce{Pb}}}\xspace}
\newcommand{\AuAu}{{\ensuremath{\ce{Au}\ce{Au}}}\xspace}

\newcommand{\DstDzPi}{\mbox{\ensuremath{\Dstarp\to\Dz(\Km \pip)\pip}}\xspace}
\newcommand{\BdJpsiKst}{\mbox{\ensuremath{\Bd \rightarrow \Kstarz(\Kp\pim) J/\psi(\mu^+ \mu^-)}}\xspace}
\newcommand{\BdToJpsiKst}{\mbox{\ensuremath{\Bd \rightarrow \Kstarz(\Kp\pim) J/\psi(\mu^+ \mu^-)}}\xspace}
\newcommand{\LzPPi}{\mbox{\ensuremath{\Lz \rightarrow p\pim}}\xspace}

\newcommand{\veloR}{\mbox{\ensuremath{{\rm \velo}\,R}}\xspace}
\newcommand{\velophi}{\mbox{\ensuremath{{\rm \velo}\,\phi}}\xspace}

\newcommand{\RunII}{Run~$2$\xspace}

\newcommand{\runtwo}{\RunII}

\newcommand{\turcall}{calibration data\xspace}

\newcommand{\nOverflowsPaper}{the number of overflows\xspace}
\newcommand{\medianDEDXPaper}{median \adc}

\newcommand{\mbias}{minimum-bias\xspace}
\newcommand{\minbias}{\mbias}
\newcommand{\ipchisq}{\chisqip}

\def\cls{CLS\xspace}

\newcommand{\NLLrich}{\ensuremath{\Lambda_{e-\pi}^{\rm RICH}}\xspace}

\usepackage{cite} 
\usepackage{mciteplus}

\usepackage{longtable} 

\begin{document}

\renewcommand{\thefootnote}{\fnsymbol{footnote}}
\setcounter{footnote}{1}



\begin{titlepage}
\pagenumbering{roman}

\vspace*{-1.5cm}
\centerline{\large EUROPEAN ORGANIZATION FOR NUCLEAR RESEARCH (CERN)}
\vspace*{1.5cm}
\noindent
\begin{tabular*}{\linewidth}{lc@{\extracolsep{\fill}}r@{\extracolsep{0pt}}}
\ifthenelse{\boolean{pdflatex}}
{\vspace*{-1.5cm}\mbox{\!\!\!\includegraphics[width=.14\textwidth]{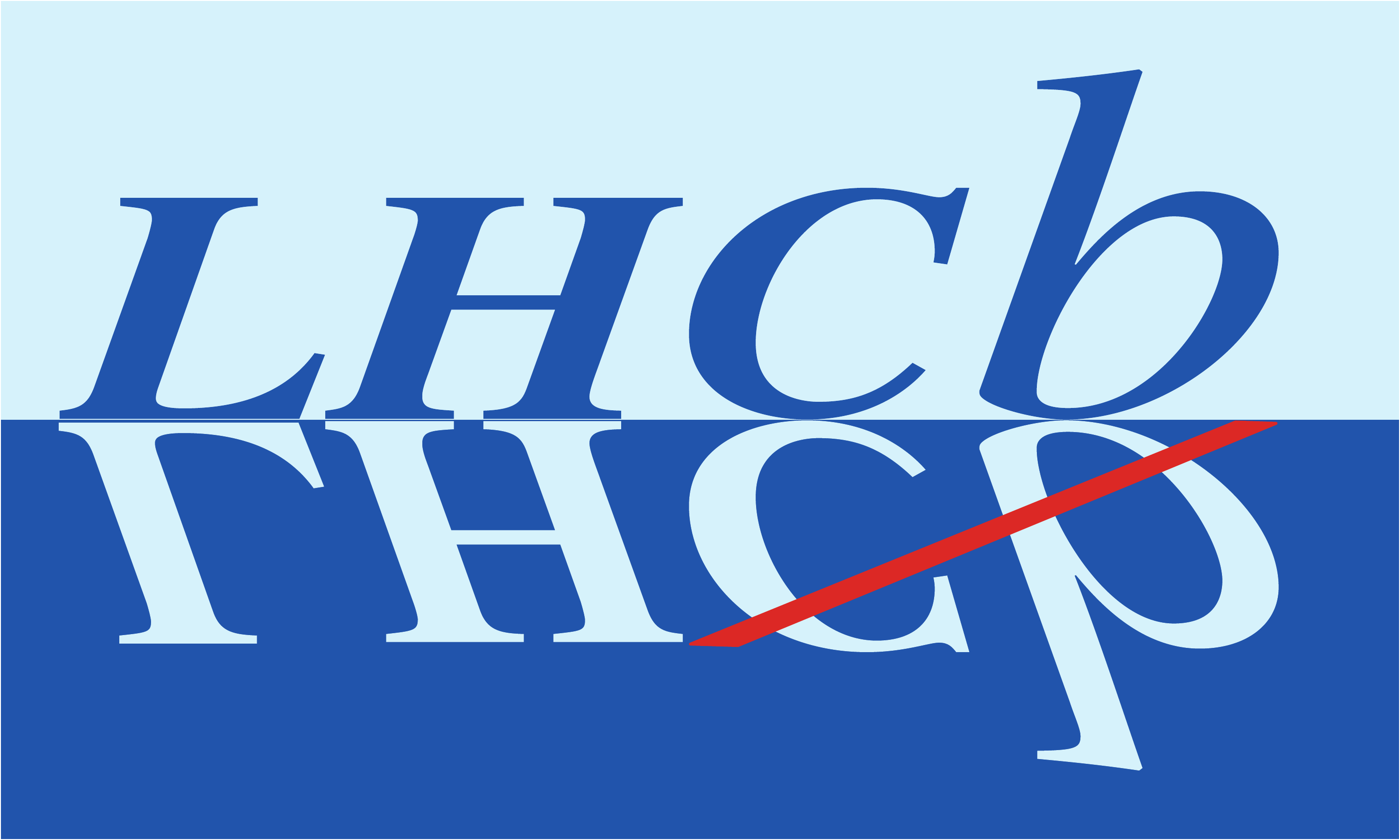}} & &}%
{\vspace*{-1.2cm}\mbox{\!\!\!\includegraphics[width=.12\textwidth]{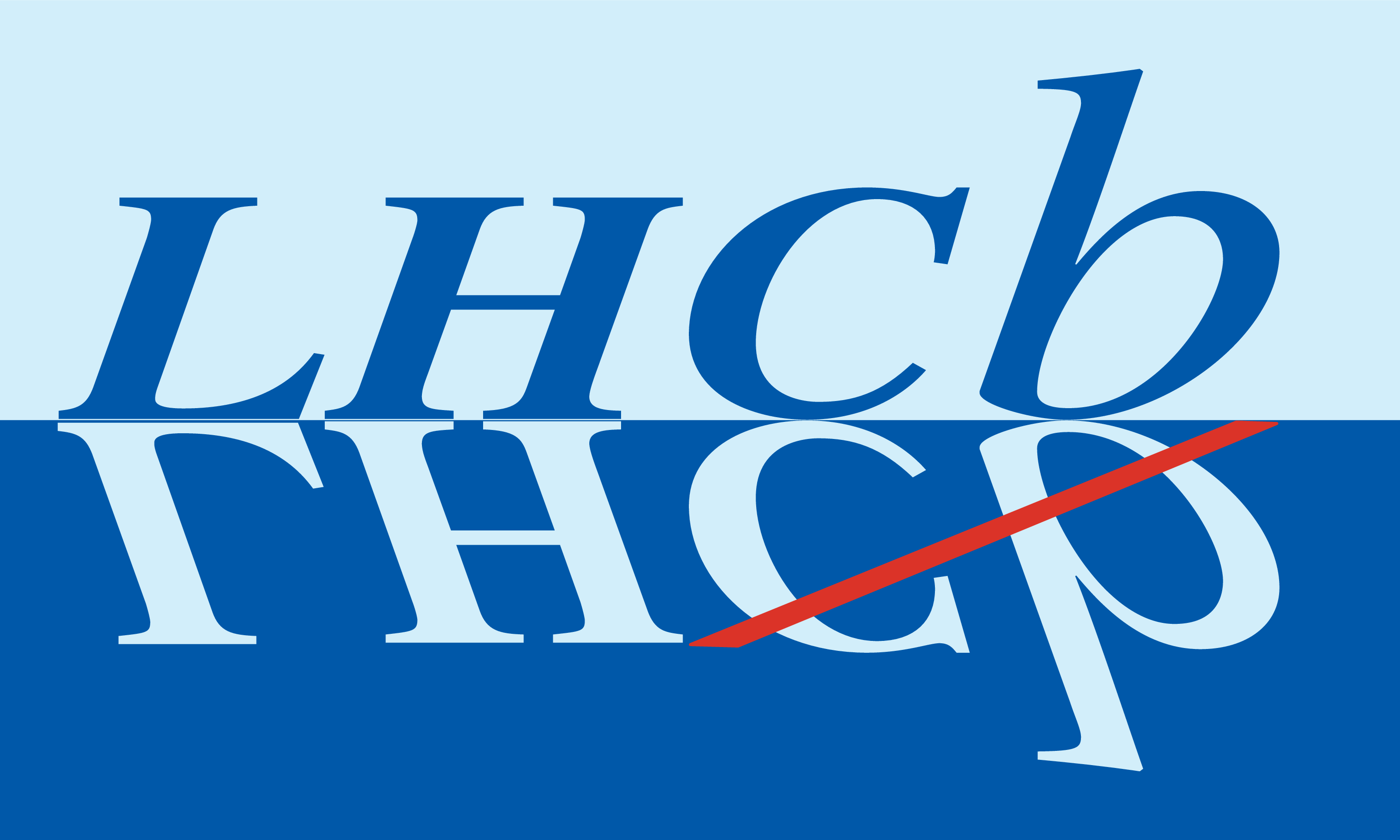}} & &}%
\\
 & & CERN-EP-2023-227 \\  
 & & LHCb-DP-2023-002 \\  
 & & $5^{\rm th}$ February 2024 \\ 
 & & \\
\end{tabular*}

\vspace*{4.0cm}

{\normalfont\bfseries\boldmath\huge
\begin{center}
  \papertitle 
\end{center}
}

\vspace*{2.0cm}

\begin{center}
\paperauthors\footnote{Authors are listed at the end of this paper.} 
\end{center}

\vspace{\fill}

\begin{abstract}
  \noindent
The identification of helium nuclei at LHCb is achieved using a method based on measurements of ionisation losses in the silicon sensors and timing measurements in the Outer Tracker drift tubes.
The background from photon conversions is reduced using the RICH detectors and an isolation requirement.
The method is developed using \pp collision data at \mbox{$\sqs=13\tev$} recorded by the LHCb experiment in the years 2016 to 2018, corresponding to an integrated luminosity of 5.5\invfb.
A total of around $10^5$ helium and antihelium candidates are identified with negligible background contamination. 
The helium identification efficiency is estimated to be approximately $50\%$ with a corresponding background rejection rate of up to $\mathcal O(10^{12})$.
These results demonstrate the feasibility of a rich programme of measurements of QCD and astrophysics interest involving light nuclei. 
  
\end{abstract}

\vspace*{2.0cm}

\begin{center}
Published in JINST 19 (2024) P02010
\end{center}

\vspace{\fill}

{\footnotesize 
\centerline{\copyright~\papercopyright. \href{\paperlicenceurl}{\paperlicence}.}}
\vspace*{2mm}

\end{titlepage}


\newpage
\setcounter{page}{2}
\mbox{~}
%
%
%
%

\renewcommand{\thefootnote}{\arabic{footnote}}
\setcounter{footnote}{0}




\pagestyle{plain} 
\setcounter{page}{1}
\pagenumbering{arabic}



\section{Introduction}

The observation of antihelium in Cosmic Rays (CR) could be a signature for physics beyond the standard model, such as dark matter annihilation \cite{Winkler:2020ltd} or the existence of antimatter domains in space\cite{PhysRevD.103.083016}. 
The AMS-02 experiment \cite{AMS:2021nhj} aboard the International Space Station (ISS) has reported several antihelium candidates of unclear origin~\cite{STingCERNseminar}.
Light nuclei can also be produced in interactions of CR with the interstellar medium via coalescence of secondary baryons, where the CR and the interstellar gas are mostly protons and around 10$\%$ helium \cite{AMS:2015azc}.
The coalescence of antibaryons into antinuclei has been studied at colliders, for example at the \lhc in \pp and \PbPb collisions (\alice~\cite{ALICE:2019vlx}), at the RHIC in \AuAu collisions (STAR~\cite{STAR:2019wjm}, PHENIX~\cite{PHENIX:2004vqi}, BRAHMS~\cite{BRAHMS:2010nkj}), and at the SPS in fixed-target experiments (NA49~\cite{Kolesnikov:2015ola}). 

The LHCb detector~\cite{LHCb-DP-2008-001,LHCb-DP-2014-002} covers the forward pseudorapidity region ($2<\eta<5$), as opposed to other experiments that have helium identification capabilities only at central pseudorapidity (small $|\eta|$).
The LHCb experiment can therefore measure helium production in a region that is unexplored by other experiments.
It is a single-arm forward spectrometer that includes a high-precision charged-particle reconstruction (tracking) system consisting of: a silicon-strip vertex detector~(\velo)~\cite{LHCb-TDR-005,LHCb-DP-2014-001} that surrounds the $pp$ interaction region, a large-area silicon-strip detector~(TT)~\cite{LHCb-TDR-009} located upstream of a dipole magnet with a bending power of about $4{\mathrm{\,Tm}}$, 
and three stations of silicon-strip detectors~(IT)~\cite{LHCb-DP-2013-003,LHCb-TDR-008} and straw drift tubes~(OT)~\cite{LHCb-TDR-006,LHCb-DP-2017-001} placed downstream of the magnet.
The tracking system measures the momentum \ptot of charged particles with a relative uncertainty that ranges from 0.5\% at low momentum\footnote{Natural units where $\hbar=c=1$ are used throughout.} to 1.0\% at 200\gev.
The minimum distance of a track to a primary vertex (PV), the impact parameter (IP), is measured with a resolution of $(15+29/\pt)\mum$ where \pt is the component of the momentum transverse to the beam, in\,\gev.
Different types of charged hadrons are distinguished from one another using information from two ring-imaging Cherenkov (\rich) detectors~\cite{LHCb-DP-2012-003,Calabrese:2022eju}. 

The method presented in this paper exploits information from energy losses through ionisation in the silicon sensors, alongside information from the \ot and \rich to improve the separation power between helium and minimally-ionising particles of charge $Z=1$.
It is assumed throughout this paper that all observed helium is \He, due to the suppression by a factor $\mathcal O(10^3)$ predicted by coalescence for each additional baryon~\cite{ALICE:2017xrp}. A dedicated study for the \Hefour production in \pp collisions exploring the  particle identification capabilities of the LHCb \rich is foreseen in the future.

\section{Energy-loss measurements}

\begin{figure}[!htbp]\centering
 \includegraphics[width=.49\textwidth]{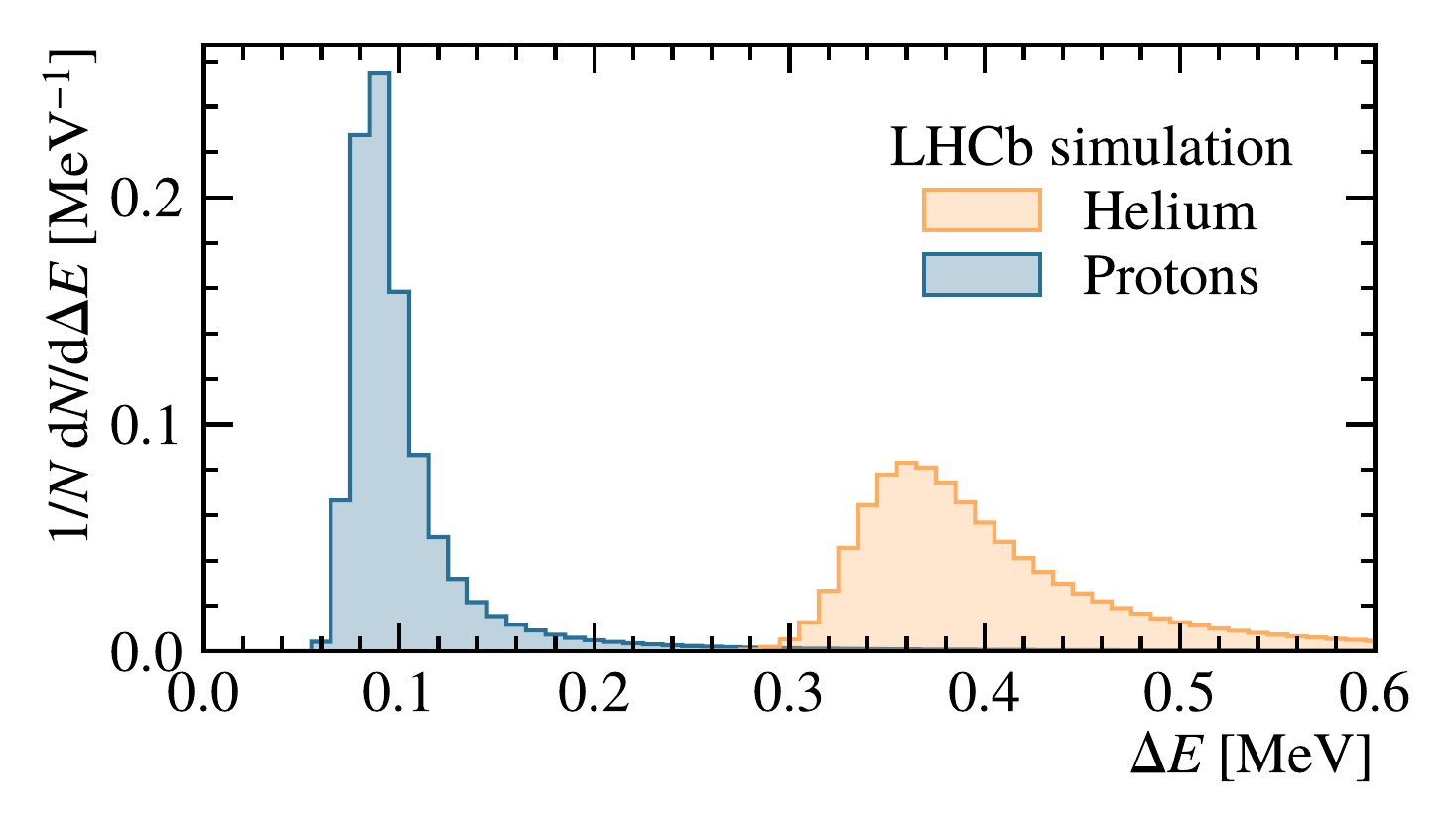} 
 \includegraphics[width=.49\textwidth]{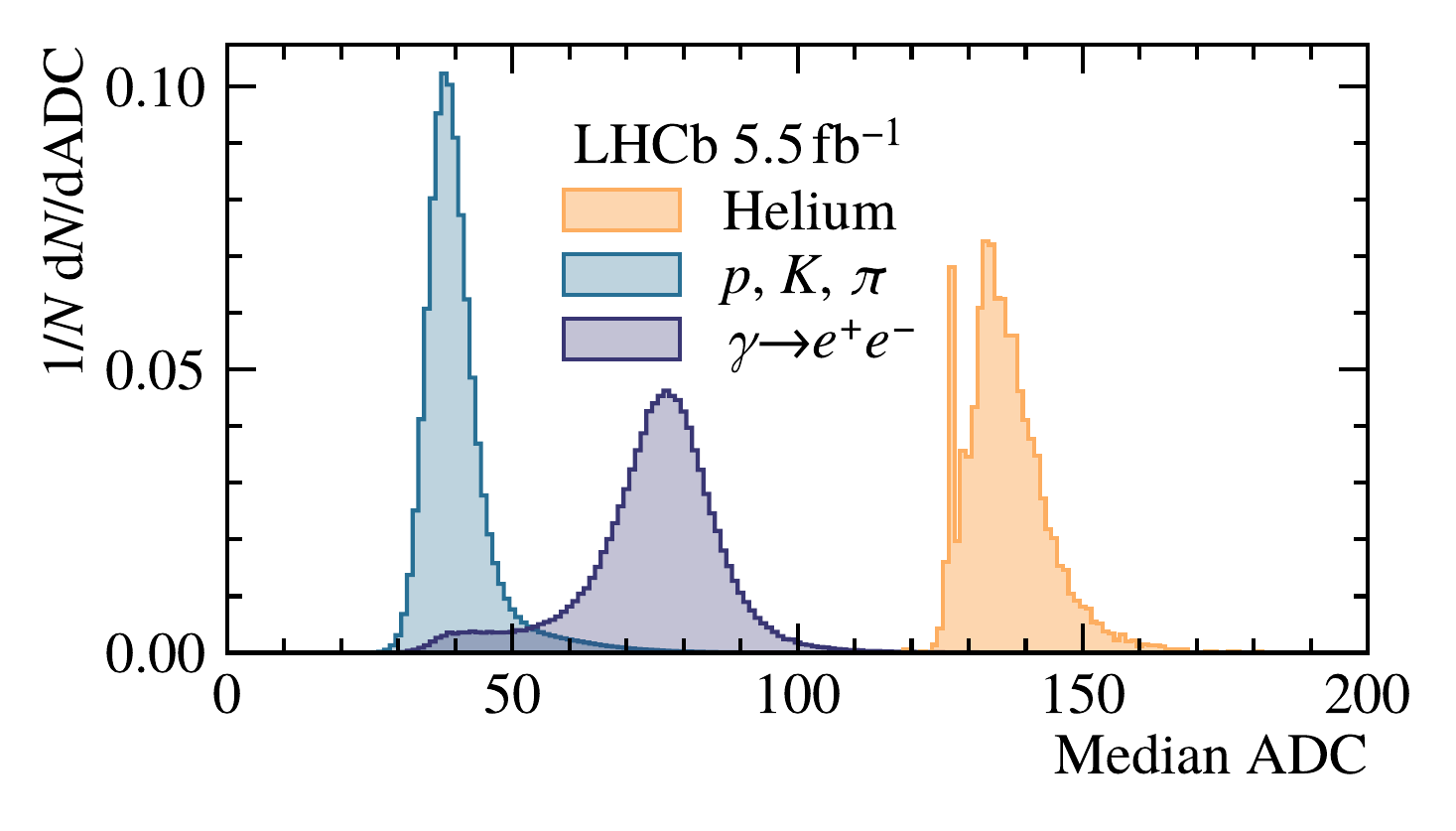} 
 \caption{Left: Distributions of the deposited energy in a \velo sensor for simulated helium and protons.
Right: Distributions of the median ADC of all \velo clusters per track, corrected for the incidence angle, for $Z=1$ particles (blue), {\ep\en pairs} from photon conversions (purple), and helium (orange).
The helium selection, described in the text, is applied but no \velo requirements are imposed.
 }\label{fig:pGun_GEANTvsADC}
\end{figure}

\noindent The average energy loss per path length of a particle of charge $Z$ passing through a silicon sensor (\dedx) is proportional to $Z^2$, as modelled by the Bethe-Bloch formula~\cite{PDG2022}.
{Figure~\ref{fig:pGun_GEANTvsADC} illustrates the energy deposited by helium and $Z=1$ particles}. 
In silicon, these energy losses are converted into around 80 electron-hole pairs per micron, and the movement of this charge in the electric field of the depleted region induces a signal on the readout strips.
This signal can appear in one or a few strips, depending on the electric field, incidence angle, and couplings between strips. 
The strips in \lhcb silicon sensors have different pitch, width, length and shapes, thus leading to different electric fields, total capacitances, and couplings.
The signal from each strip is registered by the front-end electronics and measured by the readout system, which is equipped with 7-bit analogue-to-digital converters. 
This limits the dynamical range to 127 counts per strip, and given that one count corresponds to $\sim440$ collected electrons, signals above $\sim56\,000$ electrons saturate.
Neighbouring signal strips are typically combined into clusters of up to four strips. 
The counts from the strips in a cluster are summed together to obtain the cluster amplitude (\adc). 
The saturated strips lead to an underestimation of the cluster \adc, an effect that is partially compensated by the increasing cluster size (\cls) for $Z=2$ particles.

The distribution of the median ADC of all clusters per track, corrected for the incidence angle, is shown on the right-hand side of~\Cref{fig:pGun_GEANTvsADC}.
The sensors are calibrated such that the signal from $Z=1$ particles peaks at \adc$\sim40$, whilst the helium signal is found to peak at \adc$\sim135$.
This is as expected from the $Z^2$ dependence, taking into account saturation effects.
One such effect is the sharp feature at \adc$=127$, which corresponds to \cls$=1$ clusters that saturate.
The distribution of electrons from photon conversions is discussed in~\Cref{sec:photons}.

\section{Sample selection}
\label{sec:samples}
The results presented in this paper are obtained with the \pp-collision data collected during the years 2016, 2017, and 2018, at a centre-of-mass energy $\sqs=13\tev$.
The total integrated luminosity of 5.5~\invfb is composed of 1.6~\invfb collected in 2016, 1.7~\invfb in 2017, and 2.2~\invfb in 2018.
The combined output of all LHCb physics trigger lines~\cite{LHCb-DP-2019-001} is used.
{Two data} subsets, called ``preselection~1'' and ``preselection~2'', are defined in a data reduction stage that takes place after the trigger and full data reconstruction.

{In addition, \minbias data are used to study the performance of the identification method. This sample consists of $1.7\times10^8$ events with $2.1\times10^9$ tracks, recorded with random triggers and selected using a small prescaling factor.}
Reference samples of singly-charged particles are also obtained from \LzPPi and \DstDzPi decays\footnote{The inclusion of charged-conjugate processes is implied throughout, unless otherwise stated.} that are processed into dedicated calibration data~\cite{Aaij:2018vrk}.
They contain around $10^9$ tracks in total and correspond to an integrated luminosity of 3\invfb.\footnote{This calibration dataset only covers 2018 and part of 2017, due to missing ADC information.}
These samples are supplemented by kaons and pions from \BdJpsiKst candidates selected from data corresponding to $5.5\invfb$. 

The response of the LHCb detector to \He is simulated using a simplistic Monte Carlo simulation where a single particle is generated at the nominal \pp interaction point and traced through the \lhcb detector using the \geant toolkit~\cite{Allison:2006ve, *Agostinelli:2002hh}.
The simulated events are processed by the same reconstruction software as for the data. An additional sample of simulated \BdJpsiKst decays is used for comparison with the matching data sample.

Selected tracks are required to have passed through the \velo, TT, and the three tracking stations downstream of the magnet.
A requirement is placed on the significance of the match between upstream and downstream track segments to suppress fake tracks and particles produced in interactions with the detector material.
The rigidity of each track, $\ptot/|Z|$, is required to be larger than $2.5\,\mathrm{GV}$ and the transverse component of the rigidity must 
be at least $0.3\,\mathrm{GV}$.
Each track must also be of good quality and have a sufficient number of hits in the silicon detectors to enable the identification techniques discussed below. 
Tracks that fulfil these requirements are considered well-reconstructed.


\begin{figure}[!htbp]\centering
 \includegraphics[width=.68\textwidth]{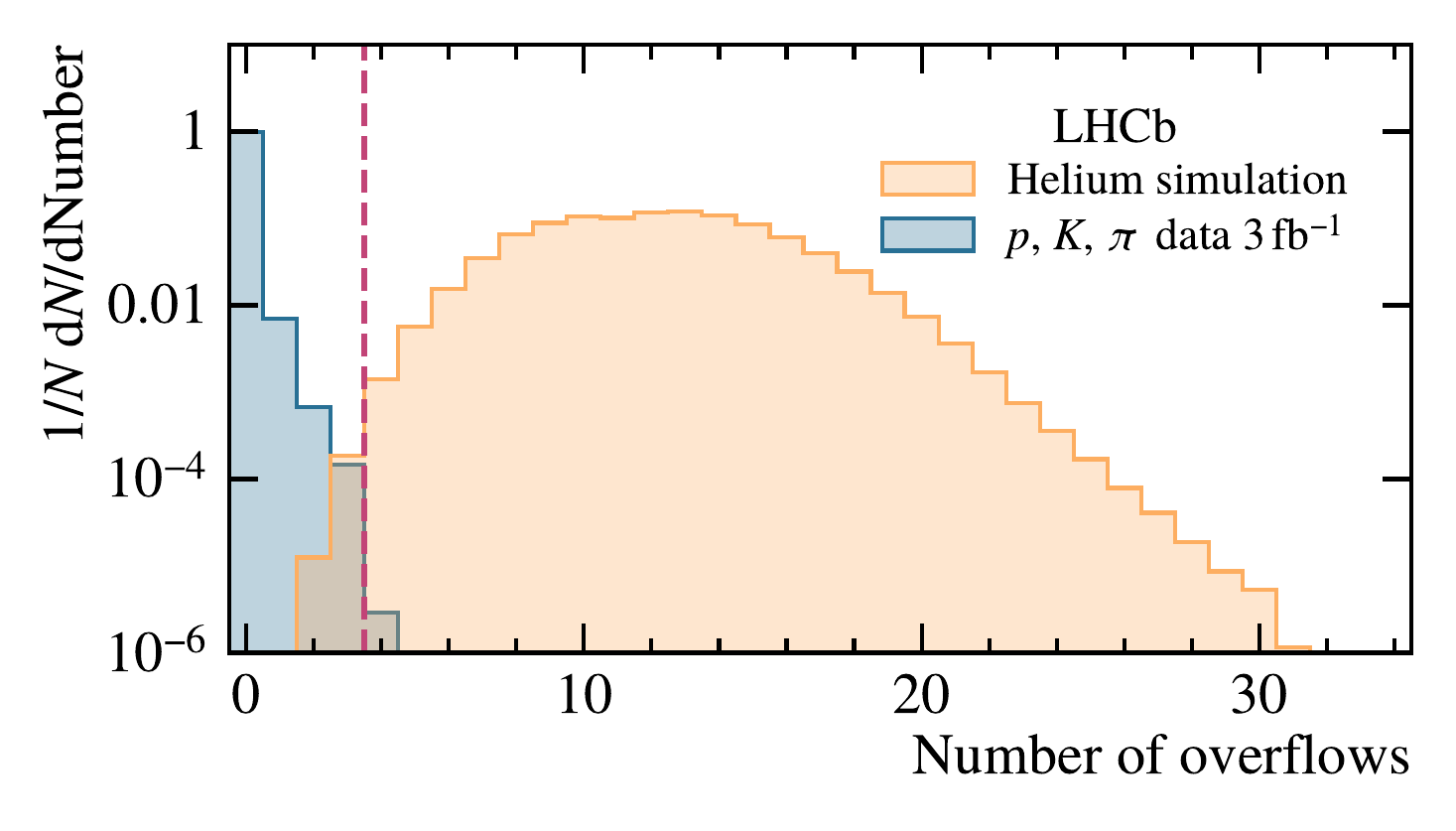} 
 \caption{Distributions of \nOverflowsPaper for simulated helium signal and $Z=1$ particles from \turcall. The vertical line indicates the preselection requirement.}\label{fig:stripping}
\end{figure}

{In the preselection stage} two simple, robust, and independent variables are used to define two subsets of data enriched in helium.
Together, they ensure redundancy in the selection, and enable efficiency comparisons.
The first subset, referred to as ``preselection~1'', requires at least one track with \nOverflowsPaper greater than 3. 
This quantity is defined as the total number of saturated strips in \velo clusters along the track that have \cls$<4$. Its distribution is shown in~\Cref{fig:stripping}, where a clear separation between simulated helium and $Z=1$ particles from \turcall is observed. The background rejection rate of this requirement is $\mathcal{O}(10^5)$.
The second subset, called ``preselection~2'', is obtained by requiring \mbox{\medianDEDXPaper$>115$}.
Additional kinematic and track-quality requirements are applied, and pions are rejected using loose particle identification cuts. 
The efficiencies of these two preselections are estimated by means described in~\Cref{sec:LLDs}.

\section{Helium identification}
The \velo, \intr, and \ttracker silicon sensors have different geometries, with the most intricate one being the \velo.
It has 42 modules (21 per side), each consisting of two different types of sensors with strips oriented to make measurements of the radial and azimuthal position. They are referred to as $R$ and $\phi$ sensors, respectively. 
The \ttracker sensors have a pitch of 183\mum and four different strip lengths, ranging from 94\mm to 378\mm. 
The \intr sensors that occupy the central part of the tracking stations downstream of the magnet have two different strip lengths, 108\mm and 216\mm, with 198\mum pitch, and two different sensor thicknesses, 320\mum and 400\mum. 
The thicker sensors produce larger signals.
To account for different sensor geometries, \veloR and \velophi sensors are considered separately, as are the thin and thick IT sensors; no significant improvement is found when considering \ttracker sensors of different strip length.

\begin{figure}[!htbp]\centering
 \includegraphics[width=0.49\textwidth]{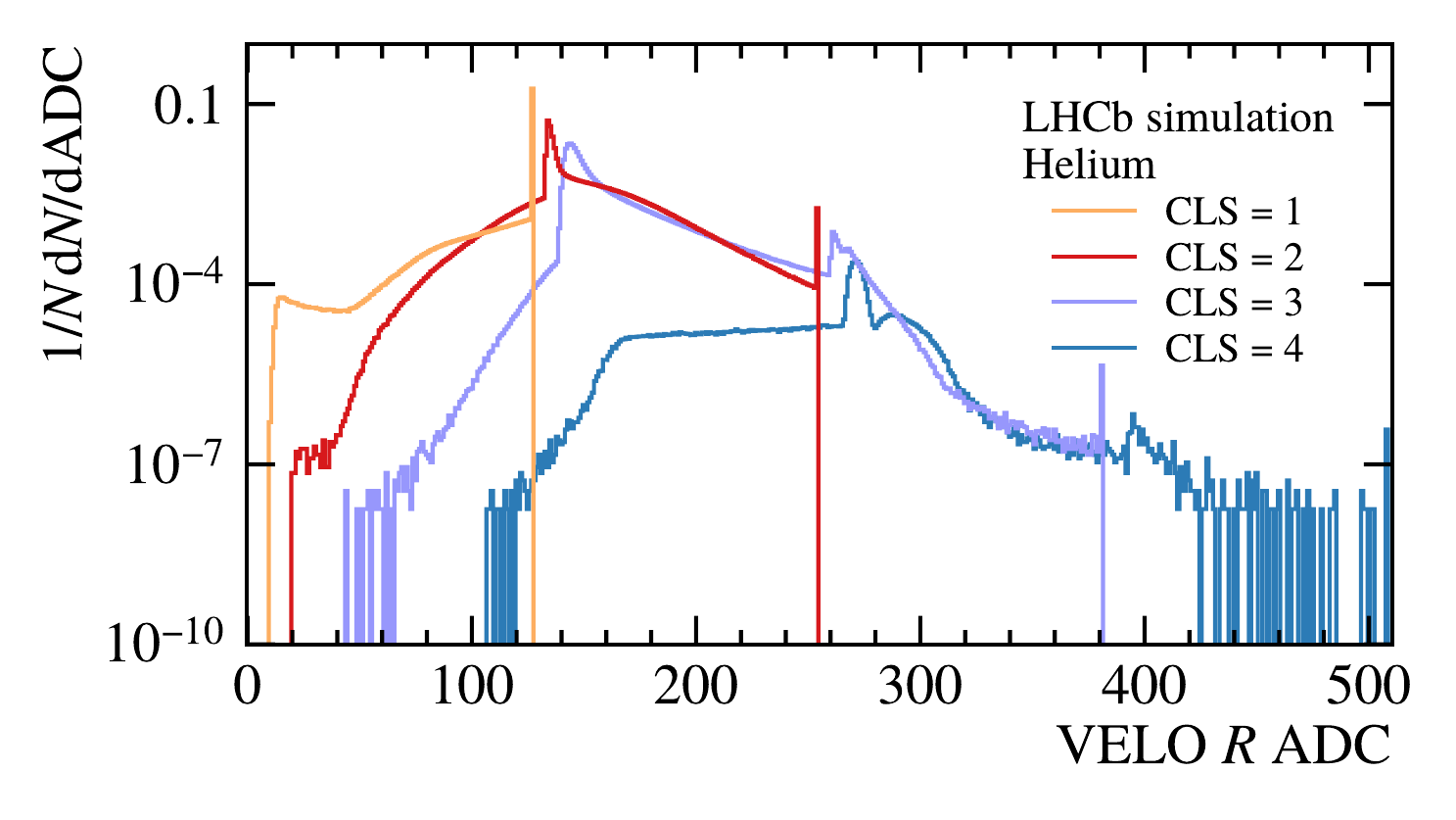} 
 \includegraphics[width=0.49\textwidth]{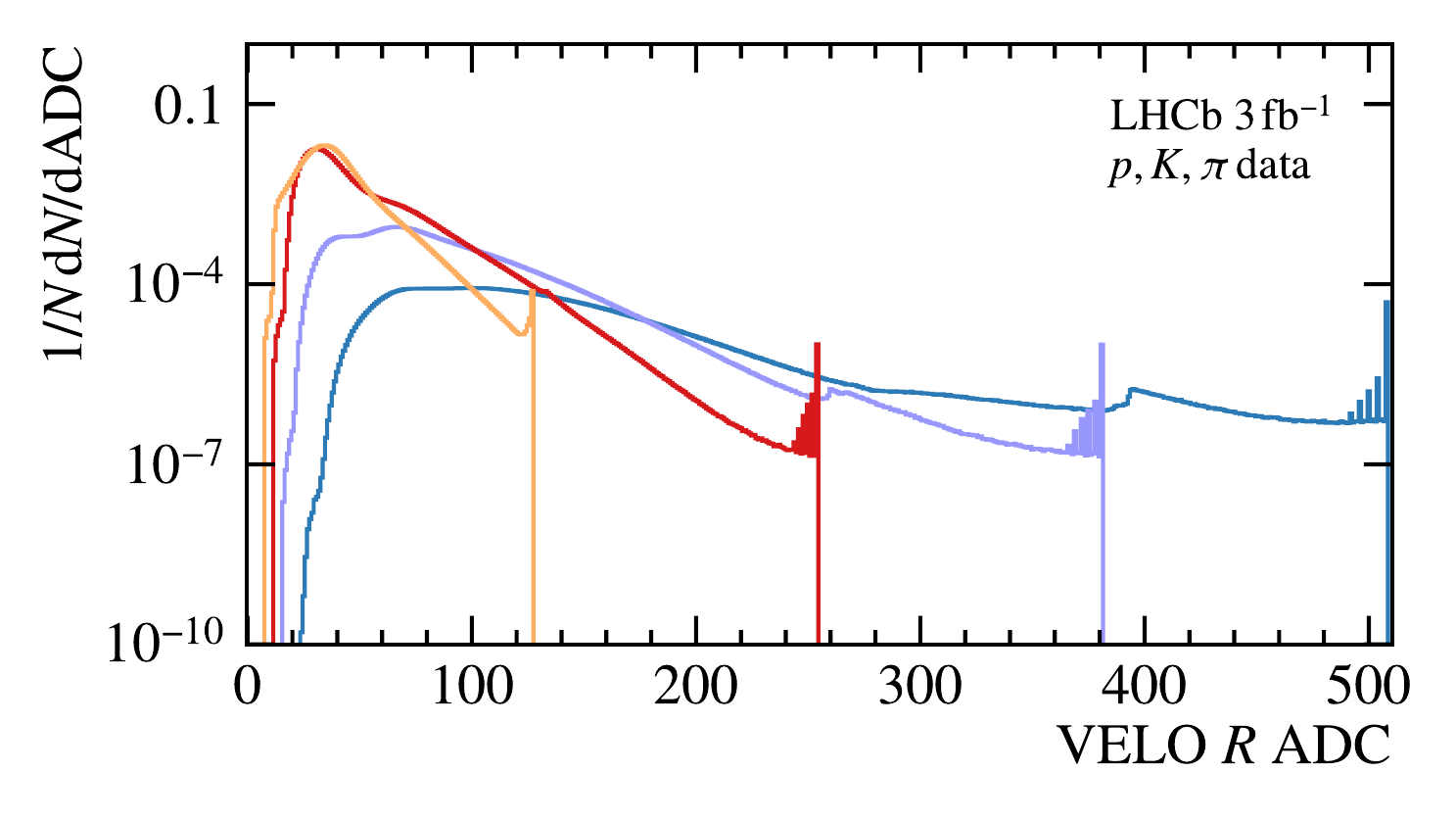} 
\caption{Distributions of the cluster amplitudes, for different cluster sizes, in the \veloR sensors for (left) simulated helium signal and (right) \turcall.}
\label{fig:pdd_adc_proj}
\end{figure}

Figure~\ref{fig:pdd_adc_proj} shows the distribution of the cluster amplitude in \veloR sensors, separately for each cluster size, as obtained from simulated helium tracks, and $Z=1$ particles from data.
It can be seen that helium is characterised by larger \cls, {higher average \adc values}, and more frequent saturation.
The sharp features at \mbox{\adc = 127, 254, 381 and 508} correspond to saturation of 1, 2, 3 and 4 strips, respectively.

\subsection{Likelihood estimators from silicon detectors}
\label{sec:LLDs}

\begin{figure}[!htbp]\centering
 \includegraphics[width=0.49\textwidth]{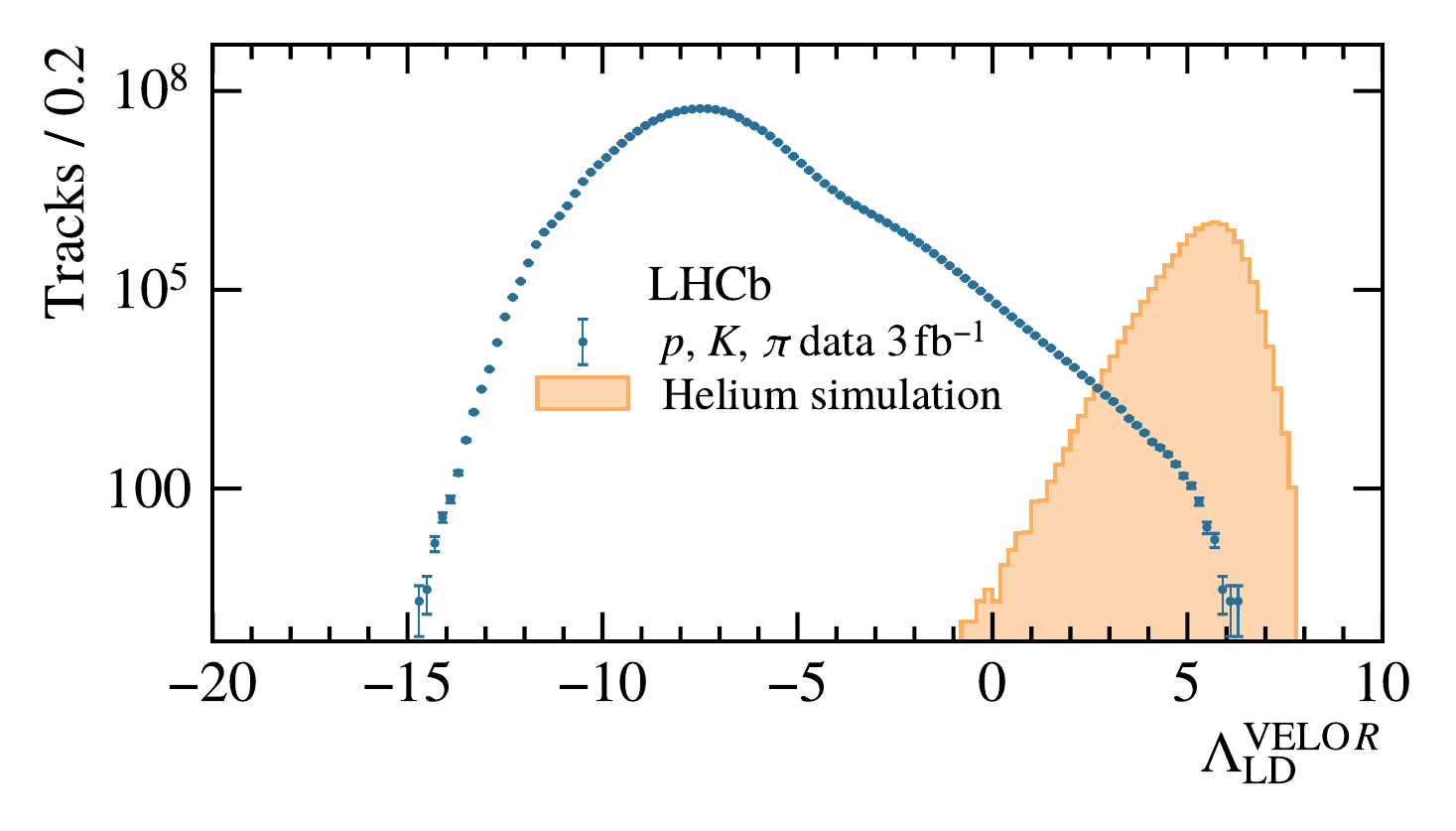}
 \includegraphics[width=0.49\textwidth]{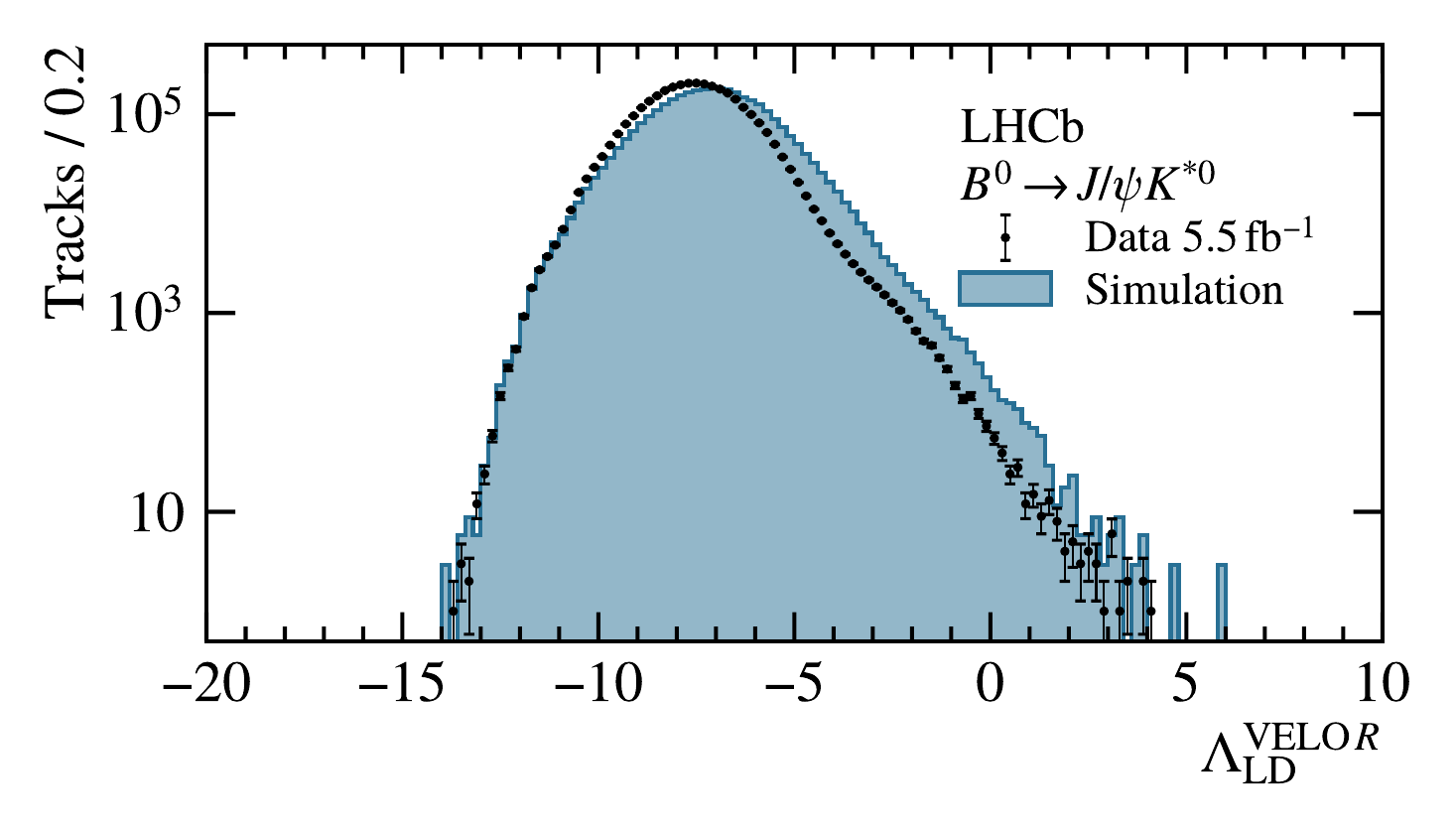}
 \caption{Distributions of \LLDveloR in: (left) samples of $Z=1$ tracks from calibration data and simulated helium; (right) samples of hadrons from \BdToJpsiKst decays. The simulation is normalised to the data.}\label{fig:llhood}
\end{figure}

Each cluster is a separate independent measurement of the energy deposit. This information is combined into a likelihood estimator for each type of sensor.
To construct this likelihood estimator, two-dimensional probability density distributions (PDD) are derived from the ADC distributions, as exemplified in~\Cref{fig:pdd_adc_proj} for \veloR sensors. 
One dimension of the PDD is the cluster size and the other is the \adc.
To derive the likelihoods of the helium {(He)} and background {(bkg)} particle hypotheses, the PDDs are used as look-up tables:
\begin{align}
\LL{X}=\left(\prod_{i=1}^{n} {\rm PDD}^X_i({\rm \cls,\adc})\right)^\frac1n,
\end{align}
where $X\in\{{\rm He},\,{\rm bkg}\}$, and $n$ is the number of clusters on the track.
Helium and background tracks are separated using the log-likelihood ratio
\begin{align}
\LLD{}=\log\frac{\LL{\rm He}}{\LL{\rm bkg}}=\log \LL{\rm He}-\log \LL{\rm bkg}.
\end{align}
The likelihoods are evaluated for \veloR, \velophi, \ttracker, \intr thick and \intr thin sensors.
The likelihood for the former is shown as an example in~\Cref{fig:llhood} on the left-hand side, where the \He simulation is compared to the calibration data described in~\Cref{sec:samples}.
Significant separation power between helium and $Z=1$ particles is observed in the likelihood estimators from all subdetectors.
However, none of them would be sufficient by themselves to produce a pure helium sample from an expected helium-to-background ratio of 1 in $\mathcal O(10^8)$.
On the right-hand side of~\Cref{fig:llhood}, the distribution of \LLDveloR in samples of pions and kaons from \BdToJpsiKst decays is compared between data and simulation.
The shape is found to be described well by the simulation; however, the tail would lead to overestimation of the background contamination in the signal region if simulation were used for such estimates.
{The differences between the distribution of helium candidates and of simulated helium are attributed to imperfect modelling of the signal in the silicon sensors.}

For the \velo, a weighted mean of the \veloR and \velophi likelihood estimators is used to define
\begin{align}
\LLDvelo=\frac{n_{{\rm \velo}\,R}\times\LLDveloR+n_{{\rm \velo}\,\phi}\times\LLDvelophi}{n_{{\rm \velo}\,R}+n_{{\rm \velo}\,\phi}}.
\end{align}
The weights $n_{{\rm \velo}\,R}$ and $n_{{\rm \velo\,\phi}}$ are the number of clusters on the track in $R$ and $\phi$ sensors, respectively.
The thin and thick \intr estimators are combined in the same way to obtain \LLDIT. 

\begin{figure}[!tbp]\centering
 \includegraphics[width=0.49\textwidth]{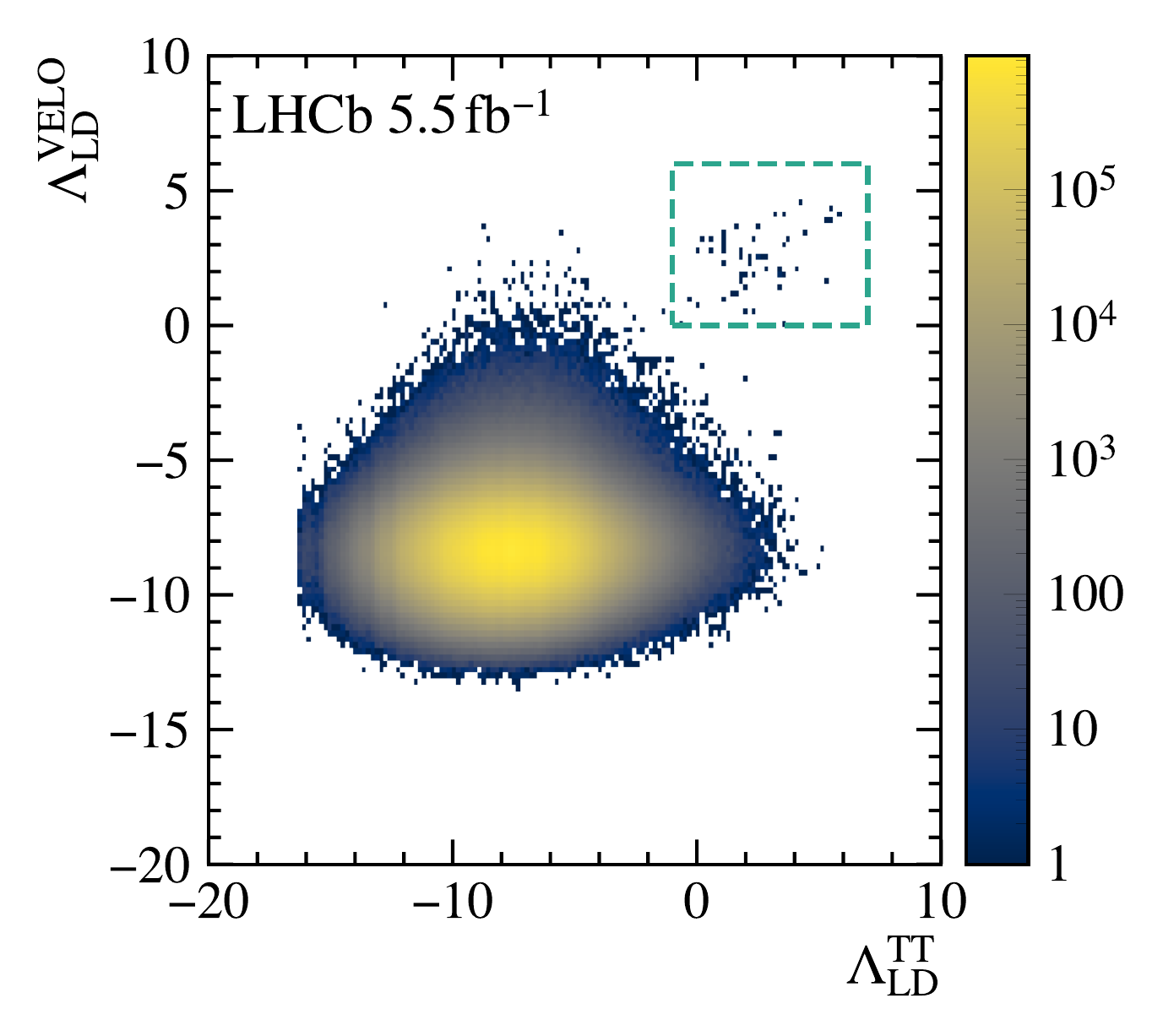}
 \includegraphics[width=0.49\textwidth]{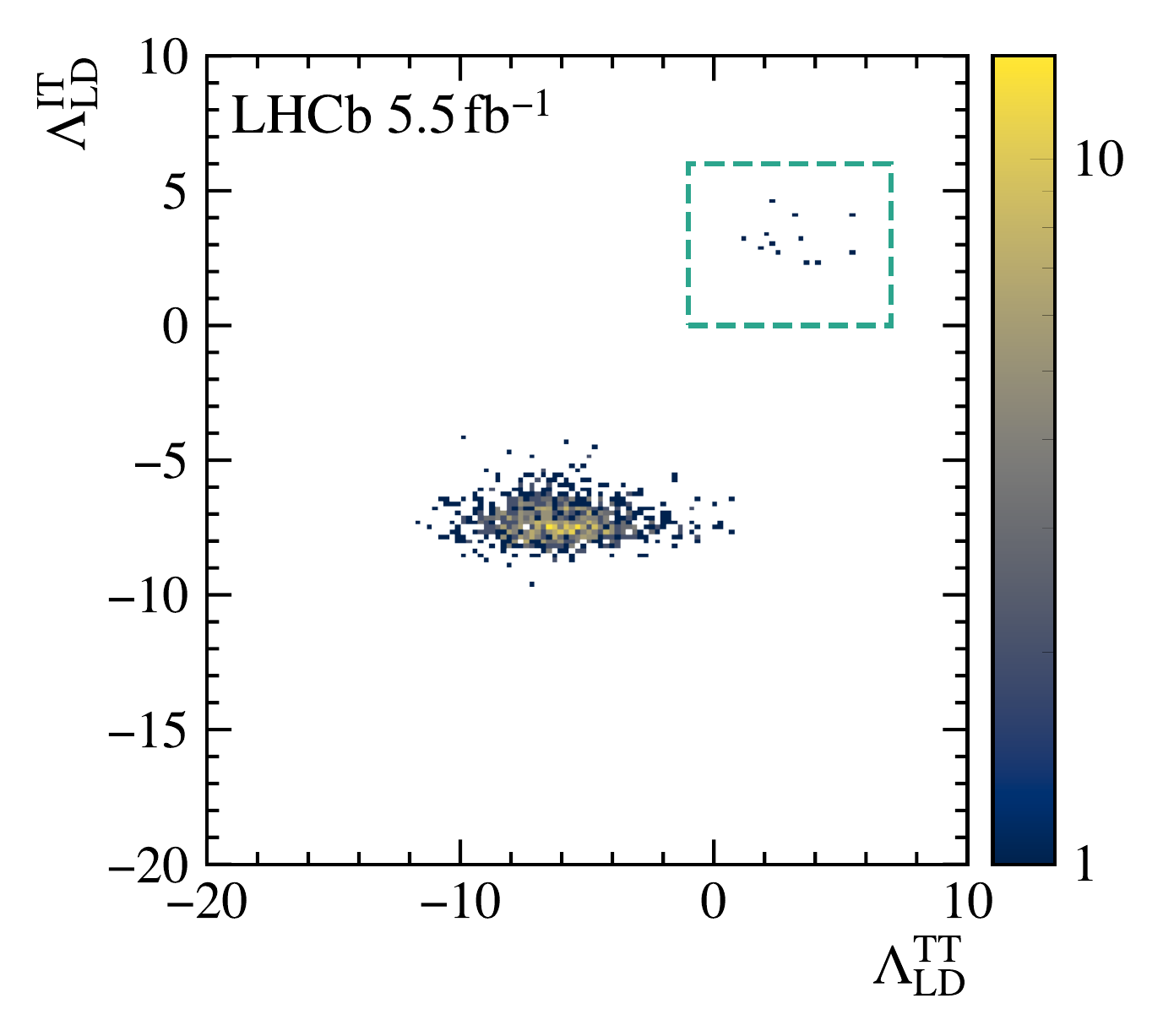}
 \caption{
Distributions of the log-likelihood estimators from the \lhcb \velo (\LLDvelo), TT (\LLDTT) and IT (\LLDIT) silicon trackers from the \minbias data.
{Helium candidates are indicated by green boxes.}
Left: the tracks are required to pass the selection described in~\Cref{sec:tracktime} or to have $\LLDIT>-1$.
Right: the tracks must traverse the IT and have $\LLDvelo>0$.
The full selection described in~\Cref{sec:tracktime,sec:photons,sec:ipchisq} is applied.}\label{fig:2DLLD}
\end{figure}

The two-dimensional distributions of \LLDvelo versus \LLDTT, and \LLDIT versus \LLDTT in \minbias data are presented in~\Cref{fig:2DLLD}.
On the left-hand side, a population of around 50 helium candidates, clearly separated from $Z=1$ particles, is observed for $\LLDvelo>0 \text{ and } \LLDTT>-1$.
Given that there are $\mathcal O(10^9)$ tracks in the \minbias data, this indicates a separation power of at least 1 in $10^8$.
Roughly $\sim20\%$ of these tracks are in the IT acceptance and~\Cref{fig:2DLLD} (right) shows that they are predominantly distributed at large \LLDIT values.

\begin{figure}[!htbp]\centering
 \includegraphics[height=.21\textheight]{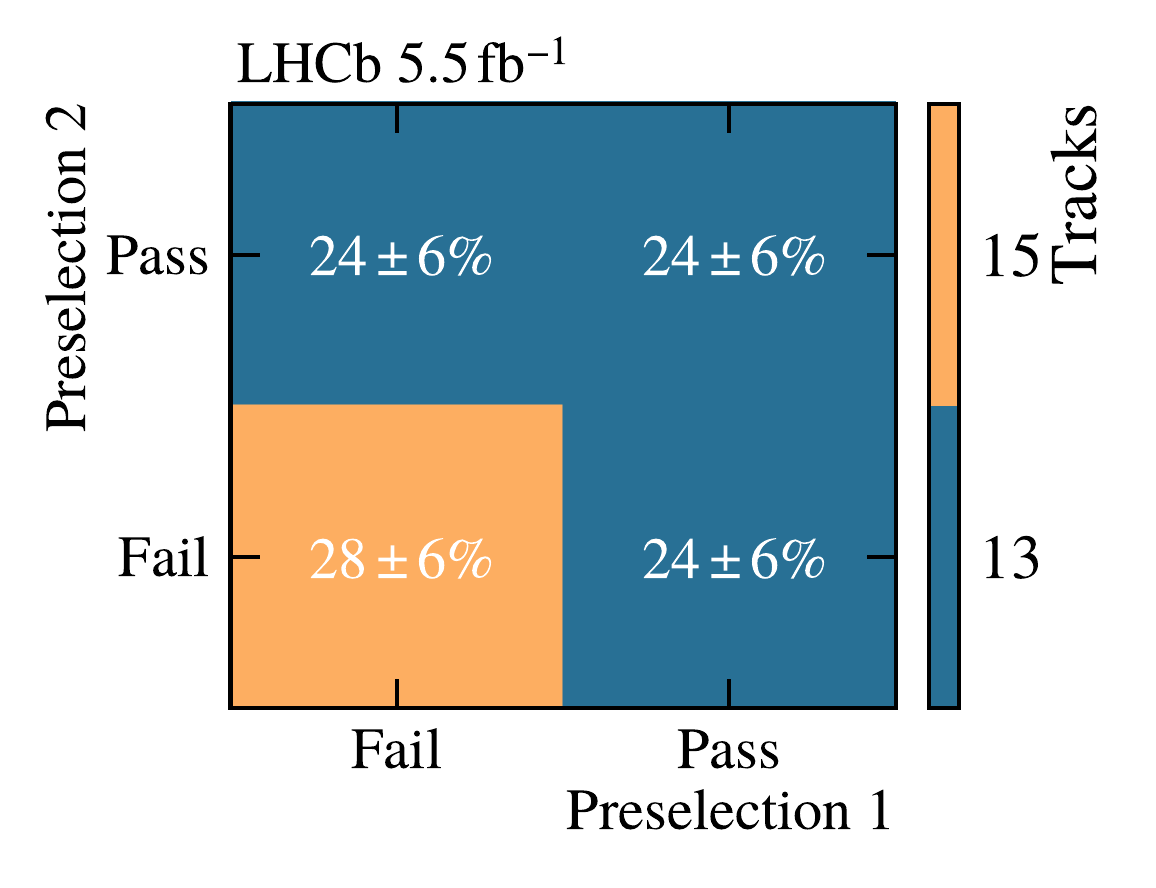} 
 \includegraphics[height=.2\textheight]{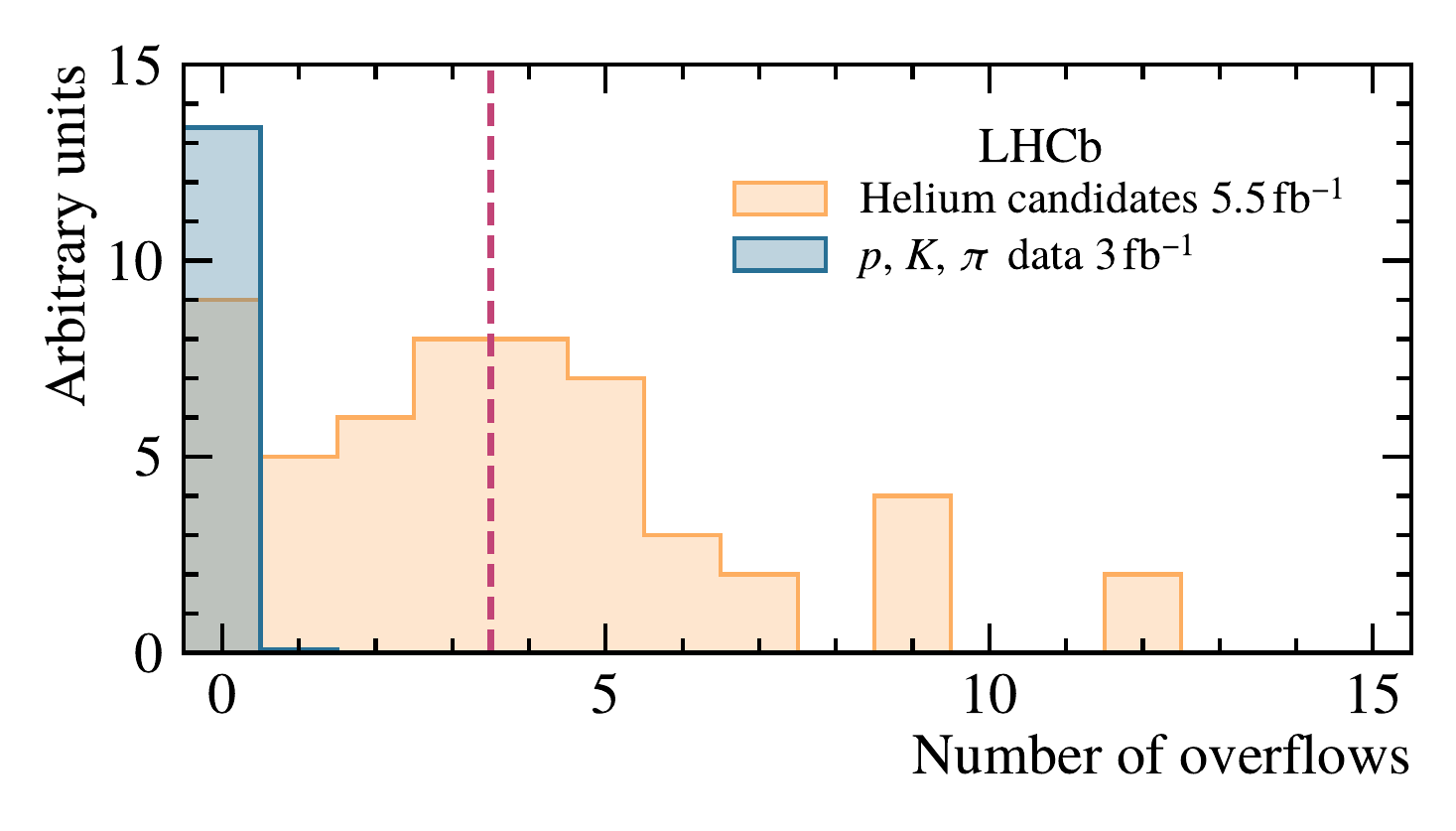} 
 \caption{Left: Efficiency matrix of the two preselections determined from the helium candidates selected in \mbias data.
 Right: Distribution of \nOverflowsPaper in helium candidate tracks from \minbias data and $Z=1$ particles from \turcall.
 The latter is normalised to 25\% of the former.
 The requirement used in the preselection is shown {by the vertical dashed line}.
 }\label{fig:minBias_stripLines}
\end{figure}

The roughly 50 helium candidates are used to estimate the efficiencies of the two preselections discussed {in~\Cref{sec:samples}}.
As shown on the left-hand side of~\Cref{fig:minBias_stripLines}, the overlap of the selected tracks is {$(24\pm6)\%$}.
In addition, both preselections are found to be approximately 50$\%$ efficient, as shown on the {right}-hand side of~\Cref{fig:minBias_stripLines} for the case of preselection~1.

\subsection{Track-time measurements from the OT} 
\label{sec:tracktime}

\begin{figure}[!htbp]\centering
\includegraphics[width=.495\textwidth]{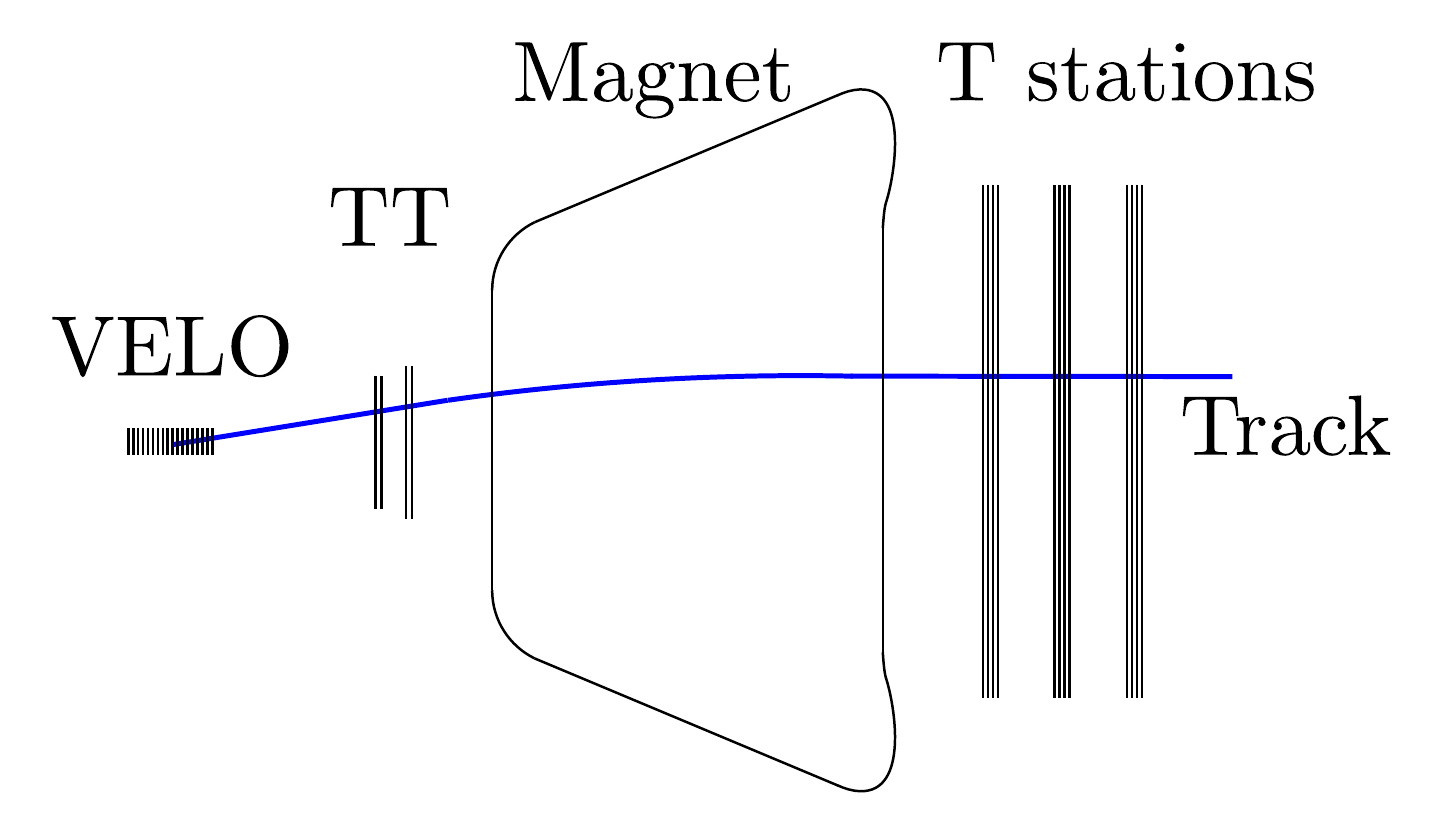} 
\includegraphics[width=.495\textwidth]{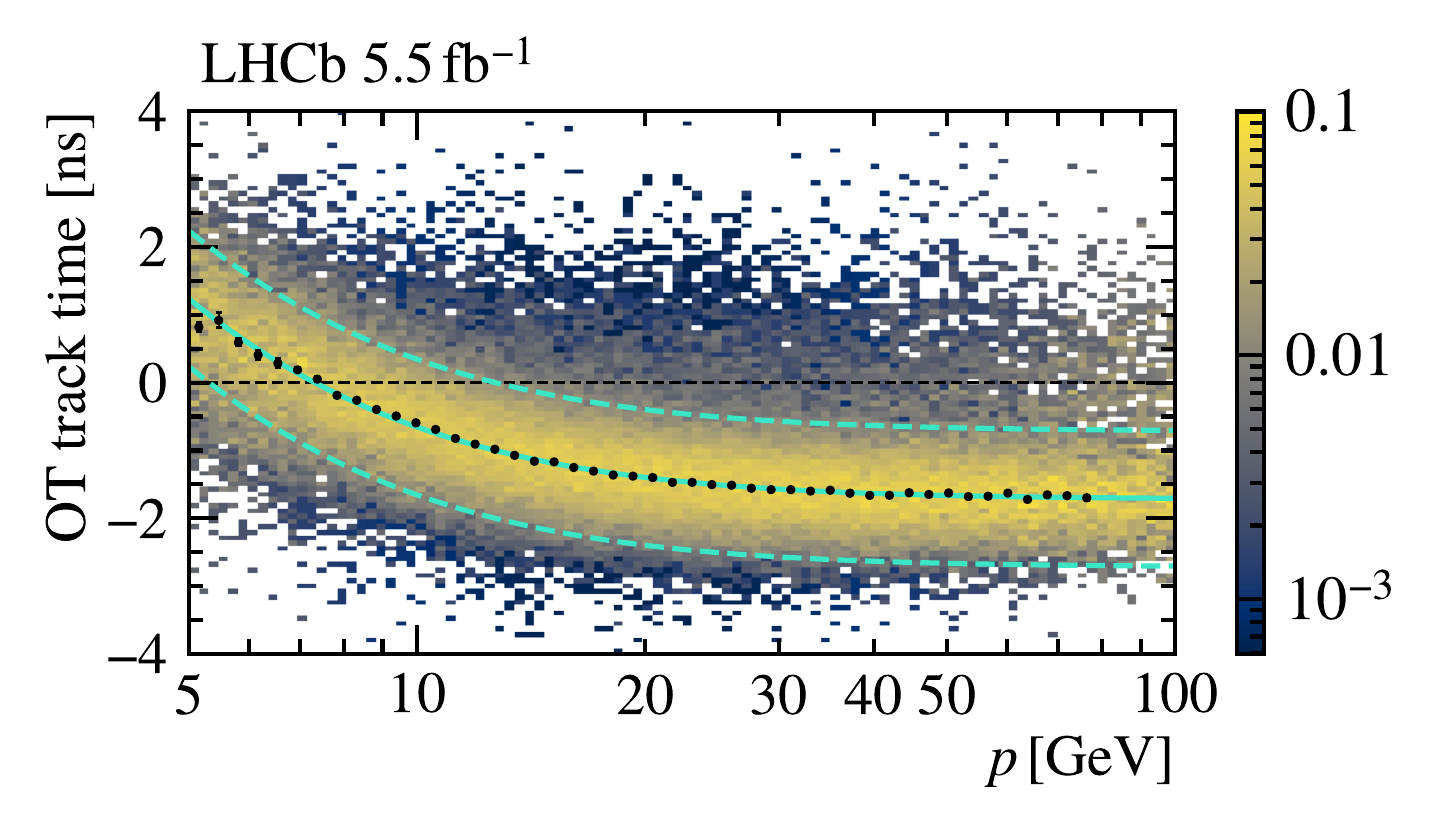} 
\caption{{Left: sketch of the LHCb tracking detectors and magnet; adapted from Ref.~\cite{LHCb-DP-2014-002}. Right:} Distribution of the reference track time versus the momentum for tracks with $\LLDvelo>0$ and $\LLDTT>1$.
In each momentum slice, the distribution is normalised to unity and fitted with a Gaussian function. The positions of the means is indicated by the black dots. These are fitted as described in the main body, resulting in the shape depicted by the green continuous line. The {dashed} green lines indicate the $\pm 1$~\ns interval.
}\label{fig:ot_fit}
\end{figure}

\noindent {As depicted in~\Cref{fig:ot_fit} on the left-hand side, }the \velo and the \ttracker stations measure the \dedx of tracks {upstream of} the magnet{. H}owever, the IT silicon detector {at the centre of the T-stations downstream of} the magnet does not cover the full detector acceptance. 
The rest of the tracks {lie in the acceptance of} the OT, which consists of gaseous drift tubes that measure the drift time of ionisation clusters produced by passing charged tracks; the time resolution is around 1\ns. 
The \ot is calibrated such that the reference track time ($t_0$) is zero for $Z=1$ particles~\cite{LHCb-DP-2017-001}.
However, helium produces on average four times more ionisation in the gas, thus triggering the readout discriminator earlier.
Therefore, helium is characterised by negative $t_0$ values at high momentum ($p>10\gev$), as shown in~\Cref{fig:ot_fit}.
Across the entire kinematic range, the shift is influenced by the momentum dependence of \dedx.
At low momentum, the time measurement is sensitive to the mass of \He, leading to an upwards shift of around 4\ns at a momentum of 5\gev.

The momentum dependence of $t_0$ is parameterised by a smooth function depicted in~\Cref{fig:ot_fit} on the right-hand side.
This function is used to calculate the expected time of a given track ($t_{\rm fit}$). 
As part of the selection of helium candidates, tracks passing through the \ot are required to have $\Delta t_{\rm \ot}\equiv |t_0-t_{\rm fit}|<1\ns$. 
This requirement significantly reduces the background with minimal signal loss.

The \velo and \ttracker estimators provide helium identification upstream of the magnet. 
The combination of the \intr and \ot provides similar information downstream of the magnet, which is essential to discriminate the signal from photon conversions, as discussed in the following section.

\clearpage
\subsection{Rejection of photon conversions} 
\label{sec:photons}

\begin{figure}[!htbp]\centering
\includegraphics[width=0.8\textwidth]{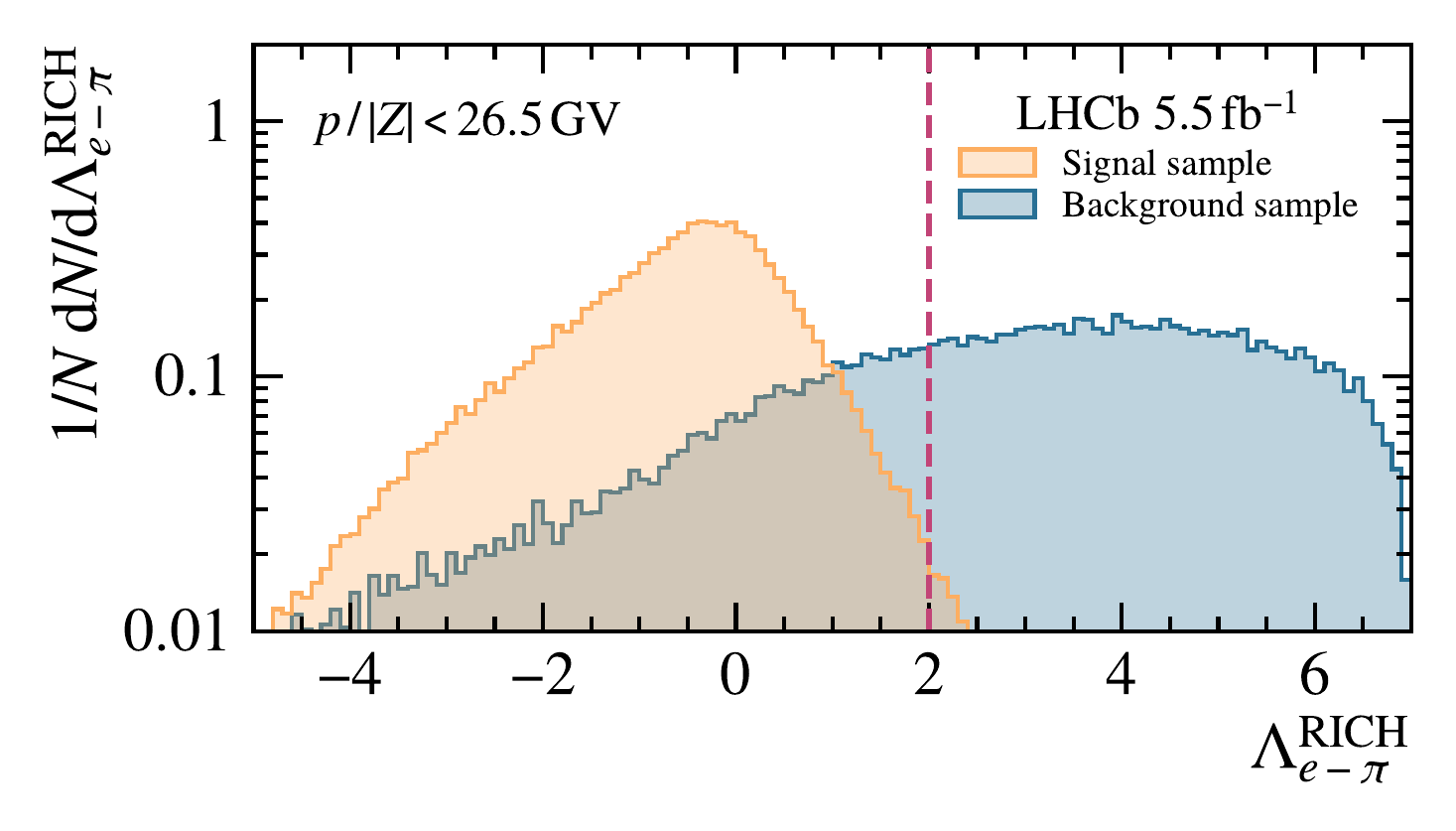} 
\caption{Distribution of the RICH electron-pion separation estimator (\NLLrich) for helium signal and background below the rigidity thresholds for Cherenkov photon production by \He in the two LHCb \rich detectors.
The vertical dashed line indicates the selection requirement.}
\label{fig:heliumRICHDLL}
\end{figure}

\noindent A photon traversing the \lhcb detector may produce an electron-positron pair with a small opening angle; the electron and positron may lose energy in the same silicon sensors, thus leading to a cluster whose amplitude is on average twice as large as expected from a $Z=1$ particle, as can be seen in the right-hand-side of~\Cref{fig:pGun_GEANTvsADC}.
Because of this, \ep\en pairs from photon conversions in the \velo are not distinguished from helium as easily as other $Z=1$ particles.
The electrons and positrons are deflected in opposite directions by the magnet but the resulting distinct tracks share the same segment in the \velo.
Therefore, conversions are suppressed by accepting only tracks that do not share their \velo segment with another track.

In addition, the \rich is used to construct log-likelihoods corresponding to different mass hypotheses for each track in the event.
These are standard variables used in \lhcb measurements, with the pion hypothesis being the default given that pions are the most produced particles.

Helium nuclei must have momenta of at least 53\gev to produce Cherenkov rings in the \rich.
Since most detected helium track candidates are below this threshold, the response of the \rich detectors to tracks in the signal sample is expected to be dominated by Cherenkov photons from other tracks in the event.
However, electrons below the pion momentum threshold may be separated using the difference between the \rich log-likelihoods constructed assuming the associated track is either a pion or an electron.
This quantity, denoted by \NLLrich, is used to reject electrons from photon conversions, as illustrated in~\Cref{fig:heliumRICHDLL}.
It depicts the distributions of \NLLrich in a helium-enriched and a background-enriched sample.
The former consists of tracks required to have $\LLDvelo>0$ and $\LLDTT>1$, whilst the latter comprises tracks selected using the cuts $\LLDvelo>0$ and $\LLDTT<-5$.
It can be seen that requiring $\NLLrich<2$ is expected to reject a substantial amount of background, whilst being highly efficient on the signal.
The separation power is diluted at high momentum, where Cherenkov rings from helium are indistinguishable from those produced by other particles.
This low-momentum background separation technique is therefore complementary to the high-momentum one provided by the OT.

\section{The helium sample}
\label{sec:results}

\begin{figure}[!htbp]\centering
\includegraphics[width=.495\textwidth]{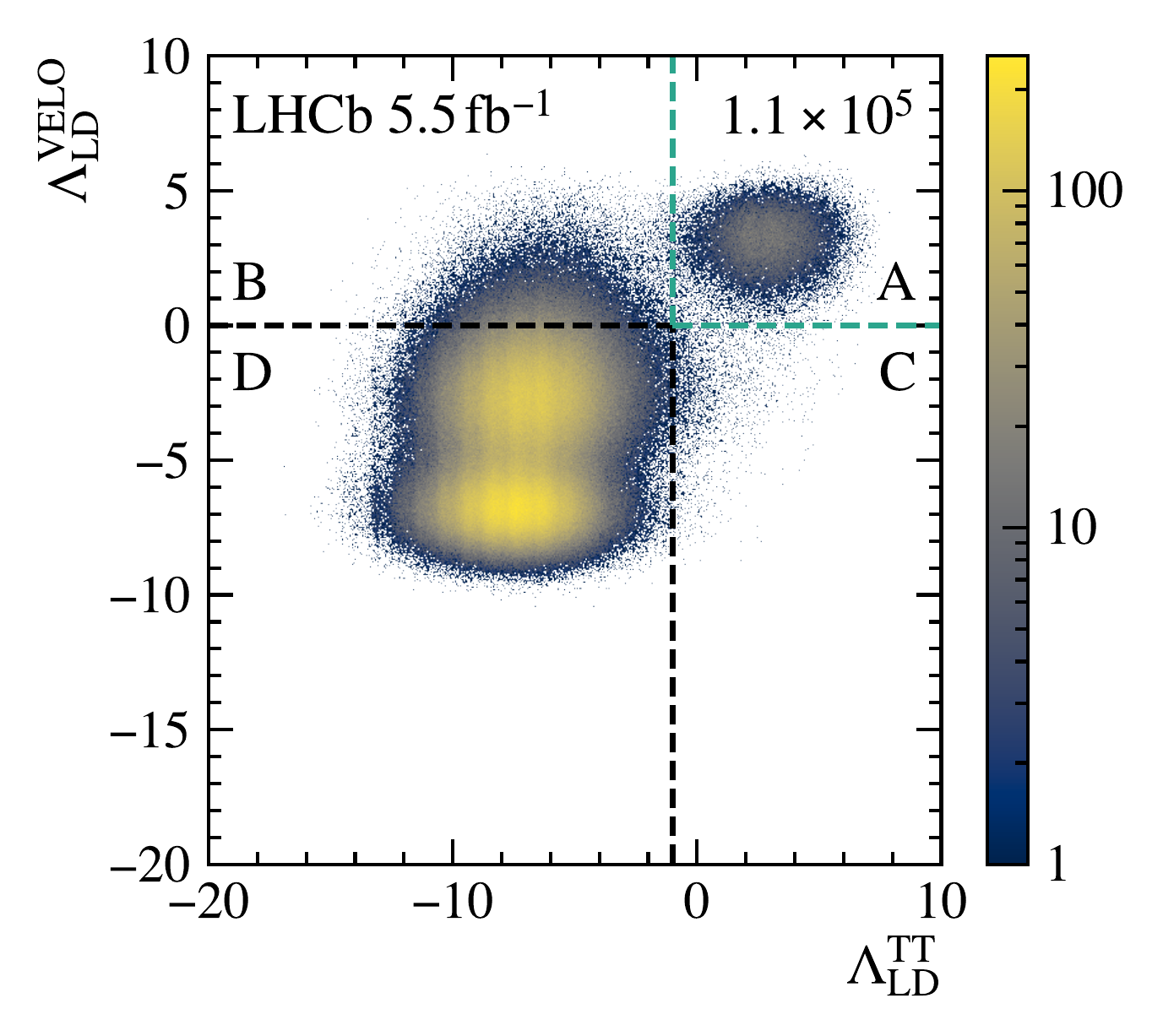}
\includegraphics[width=.495\textwidth]{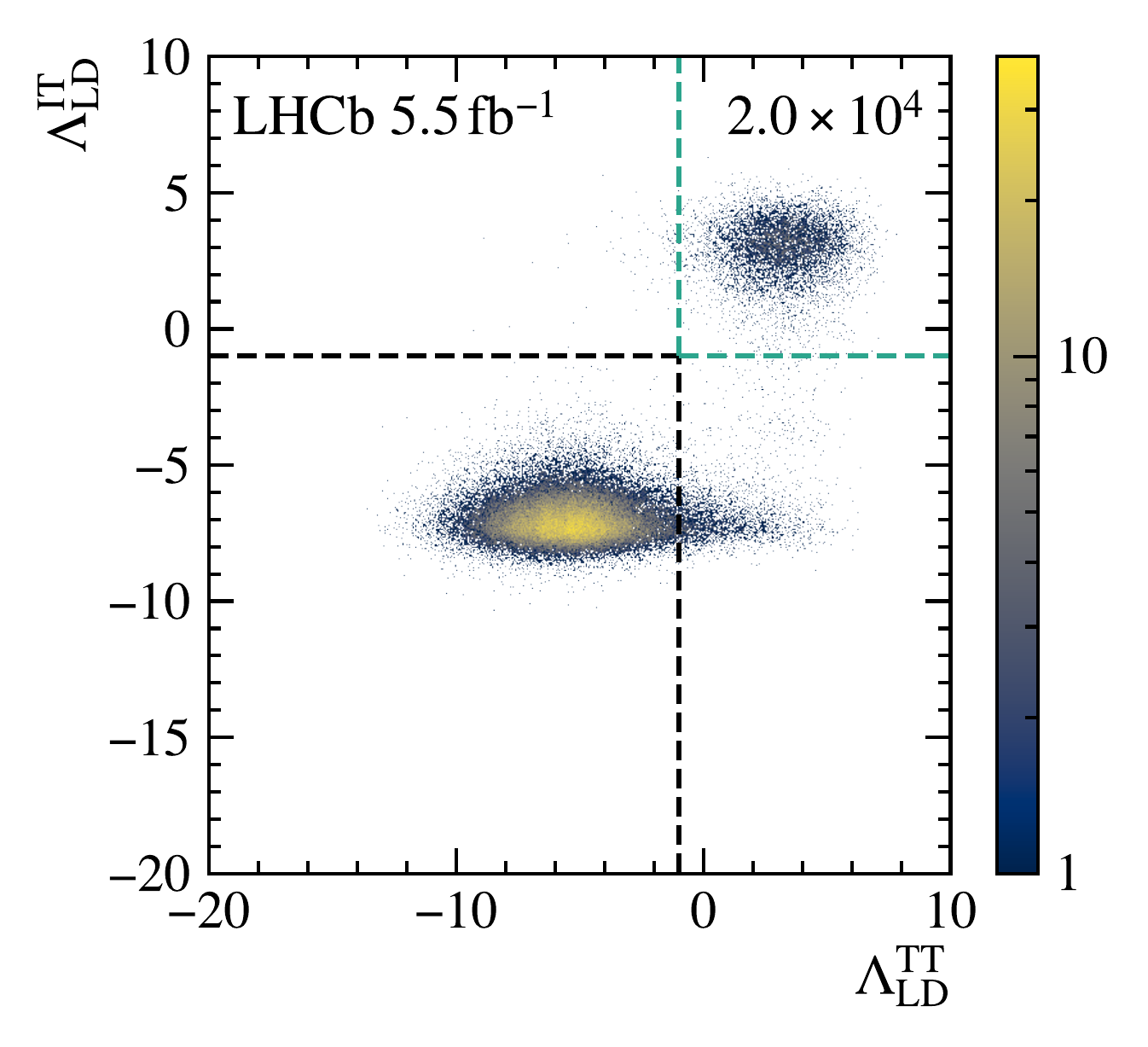} 
\caption{Distributions of the log-likelihood estimators from the \lhcb \velo (\LLDvelo), TT (\LLDTT) and IT (\LLDIT) silicon trackers. 
Left: the tracks are required to pass the downstream selection.
Right: the tracks must traverse the IT and have $\LLDvelo>0$.
On the left, the signal region is denoted by A, whilst regions B, C, and D correspond to background.
}\label{fig:LLDsABCD}
\end{figure}

\begin{table}[!htbp]
\caption{Helium selection criteria quantified throughout this paper. 
The logical or is required between the two preselection criteria.
Given the complementary acceptances of OT and IT, the logical or is also required between the downstream requirements.
}
\label{tab:selectionHelium}
\begin{center}
\begin{tabular}{l|l|r@{\,}c@{\,}r@{}l} 
\multicolumn{1}{c|}{Type}&\multicolumn{1}{c|}{Track property}&\multicolumn{4}{c}{Selection}\\ 
\hline
\multirow{3}{*}{\parbox{1.5cm}{Acceptance,\\kinematics}}&$\eta$  & \multicolumn{4}{c}{$\in(2,\,5)$}\\
&$\ptot/|Z|$ && $>$& 2.5 &\,GV\\
&$\pt/|Z|$ && $>$& 0.3 &\,GV\\
\hline
\multirow{2}{*}{Preselection}&Number of overflows && $>$& 3 &\\
&Median ADC && $>$& 115 &\\
\hline
\multirow{2}{*}{Downstream}&$\Delta t_{\rm \ot}$&& $<$&1 &$\ns$ \\ 
&\LLDIT && $>$& $-1$ \\
\hline
\multirow{2}{*}{\parbox{1.5cm}{Conversion\\rejection}}&Unique \velo segment & \multicolumn{4}{c}{Yes} \\
&\NLLrich &&$<$& 2 \\
\hline
PV origin & $\ln\ipchisq$ && $<$ & 2\\
\end{tabular}
\end{center}
\end{table}

\begin{figure}[!htbp]\centering
\includegraphics[width=.495\textwidth]{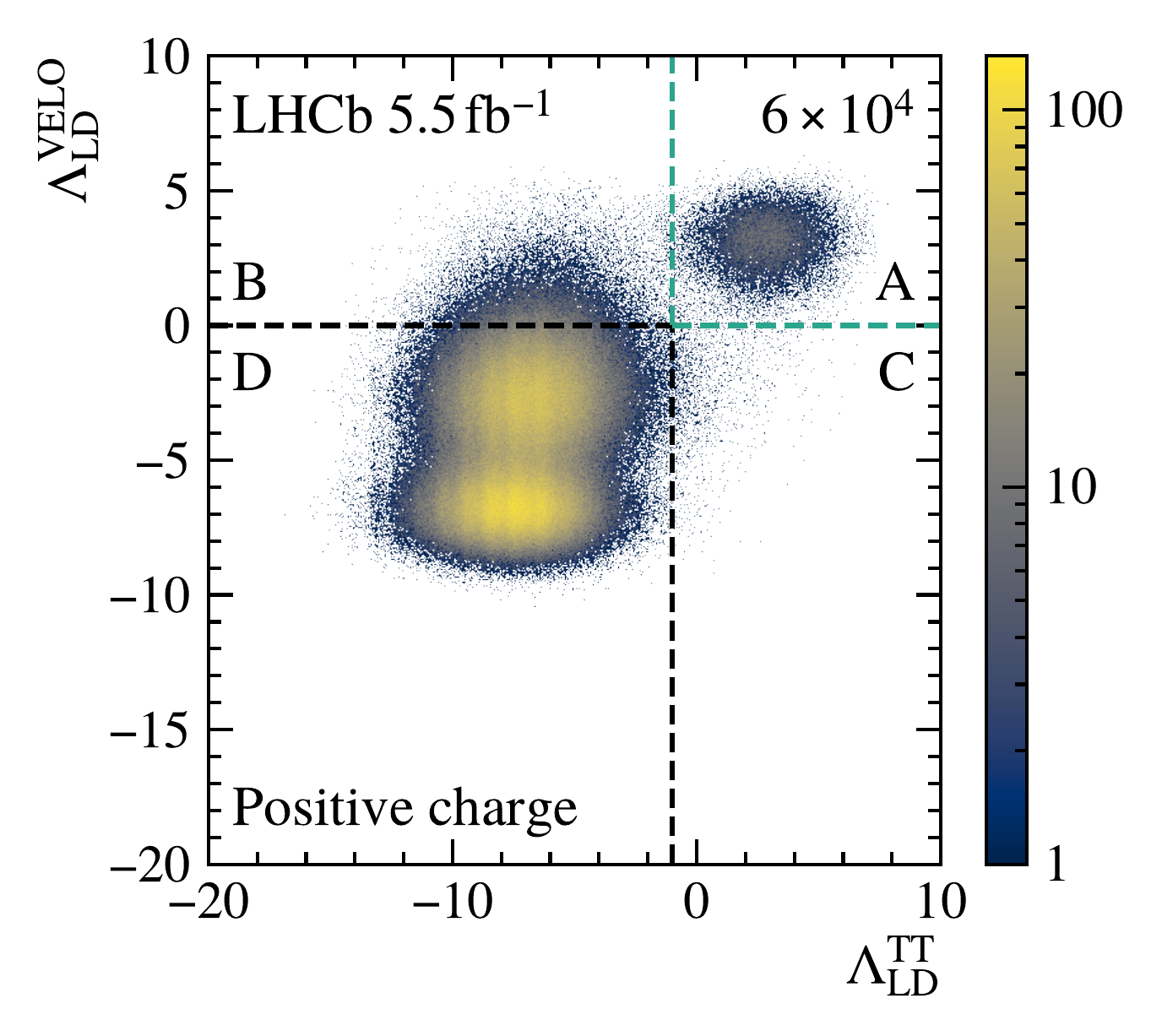} 
\includegraphics[width=.495\textwidth]{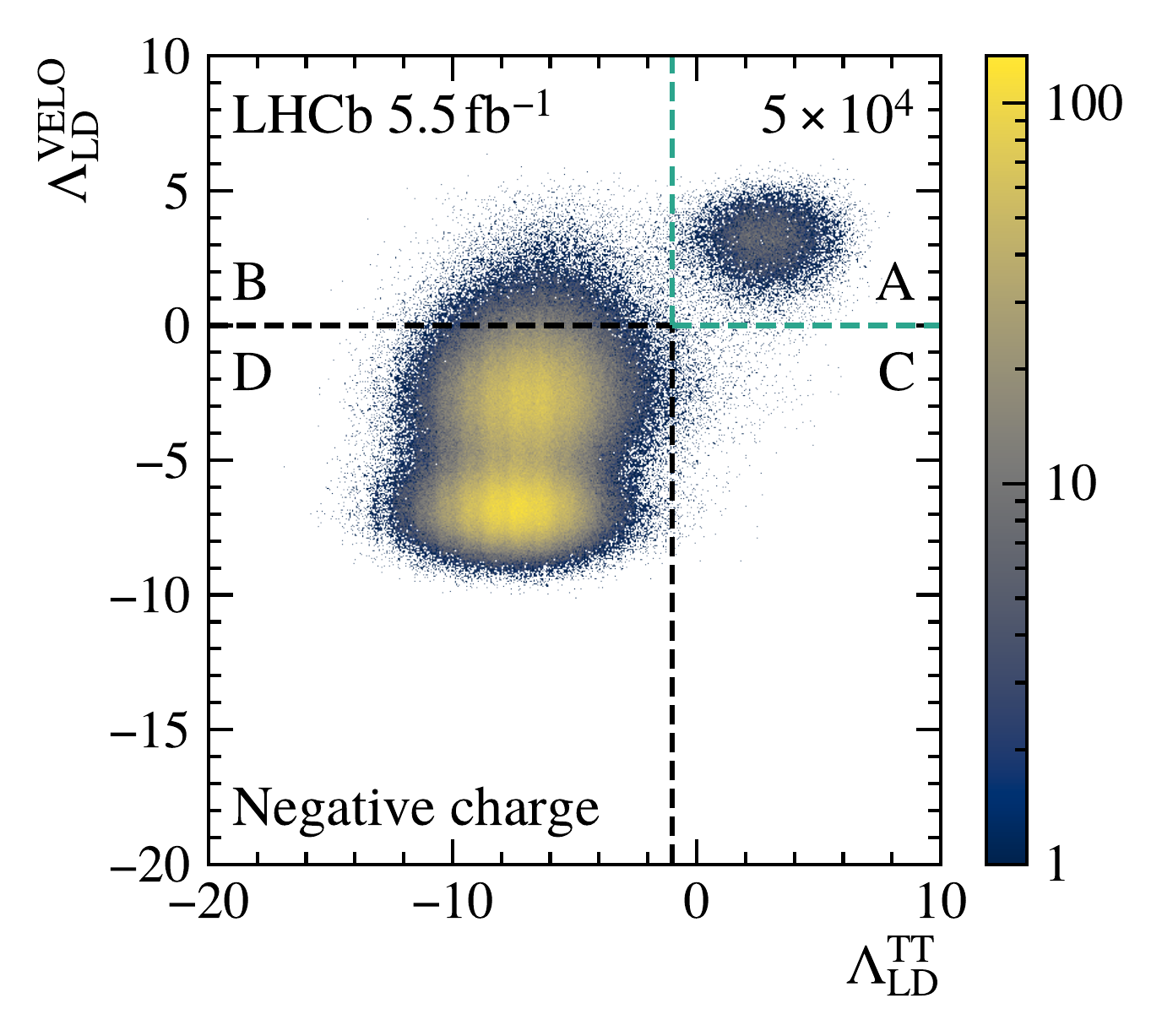} 
\caption{Distributions corresponding to that on the left-hand side of~\Cref{fig:LLDsABCD}, but separated into tracks of (left) positive and (right) negative charge.
}
\label{fig:LLDsABCD2}
\end{figure}

\noindent The data selected by applying the requirements described in the previous sections are shown in~\Cref{fig:LLDsABCD,fig:LLDsABCD2}.
Additionally, tracks that are incompatible with originating from a primary vertex are rejected by means described in the following section.
The selection criteria quantified thus far are summarised in~\Cref{tab:selectionHelium}.
A large, well-separated population of helium candidates is observed for $\LLDvelo>0$ and $\LLDTT>-1$ (region~A). 
Approximately $1.1\times10^5$ candidates are selected, and~\Cref{fig:2DLLD} shows that their location is consistent with the $\sim50$ candidates found in \minbias data. 

Singly-charged particles are located at low values of $\LLDvelo$ and $\LLDTT$ (region~D). 
They peak at two different \LLDvelo values, one for each of the preselections presented {in~\Cref{sec:samples}}.
The \medianDEDXPaper requirement of preselection~2 biases the \LLDvelo distribution towards larger values with respect to preselection~1.
Electrons from converted photons are highly collinear in the \velo but are separated by the magnetic field.
This means that they cannot be separated by \LLDvelo alone, however they can be separated by \LLDTT and are therefore found predominantly in region~B.

\begin{figure}[!htbp]\centering
\includegraphics[width=.495\textwidth]{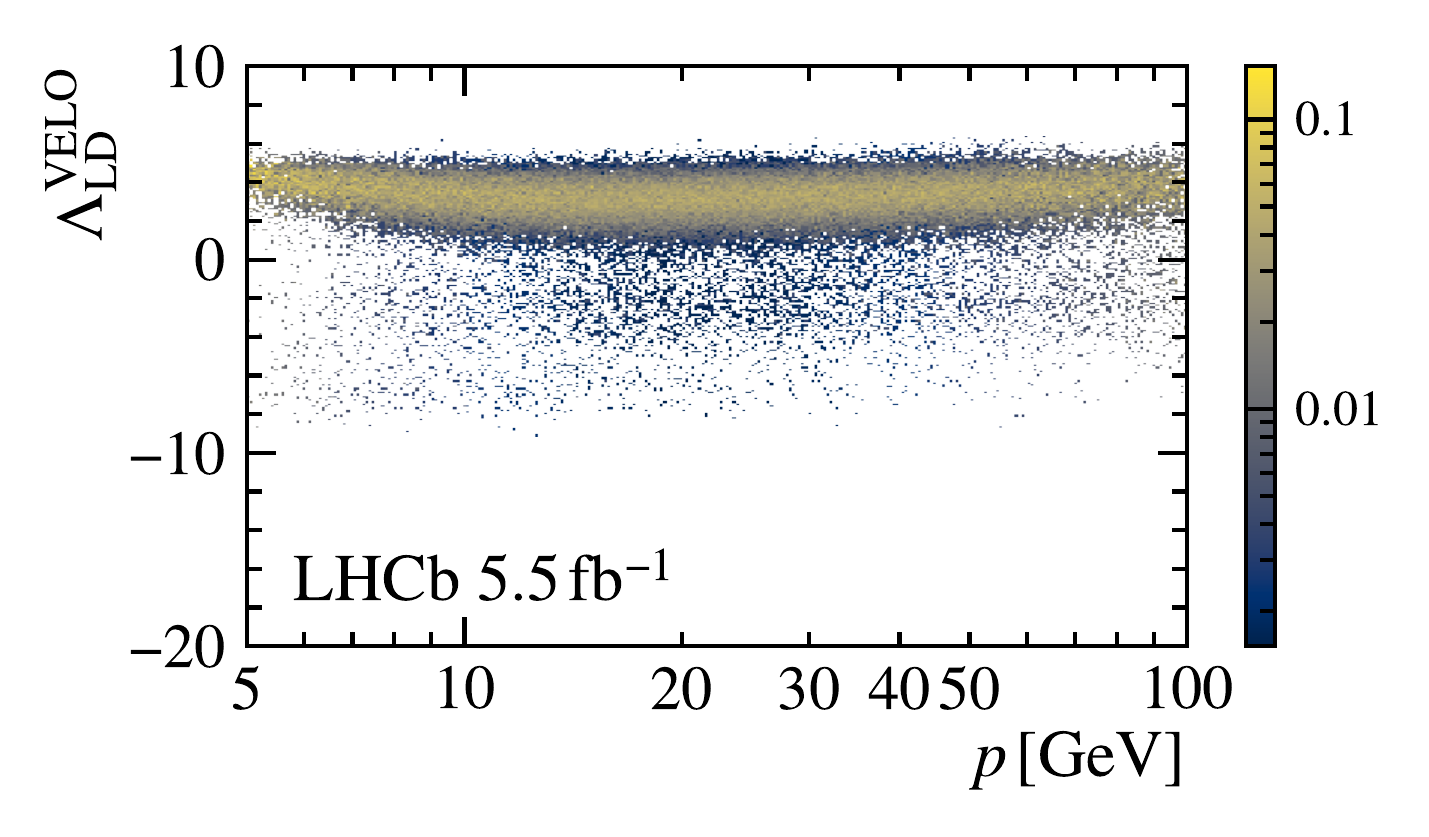} 
\includegraphics[width=.495\textwidth]{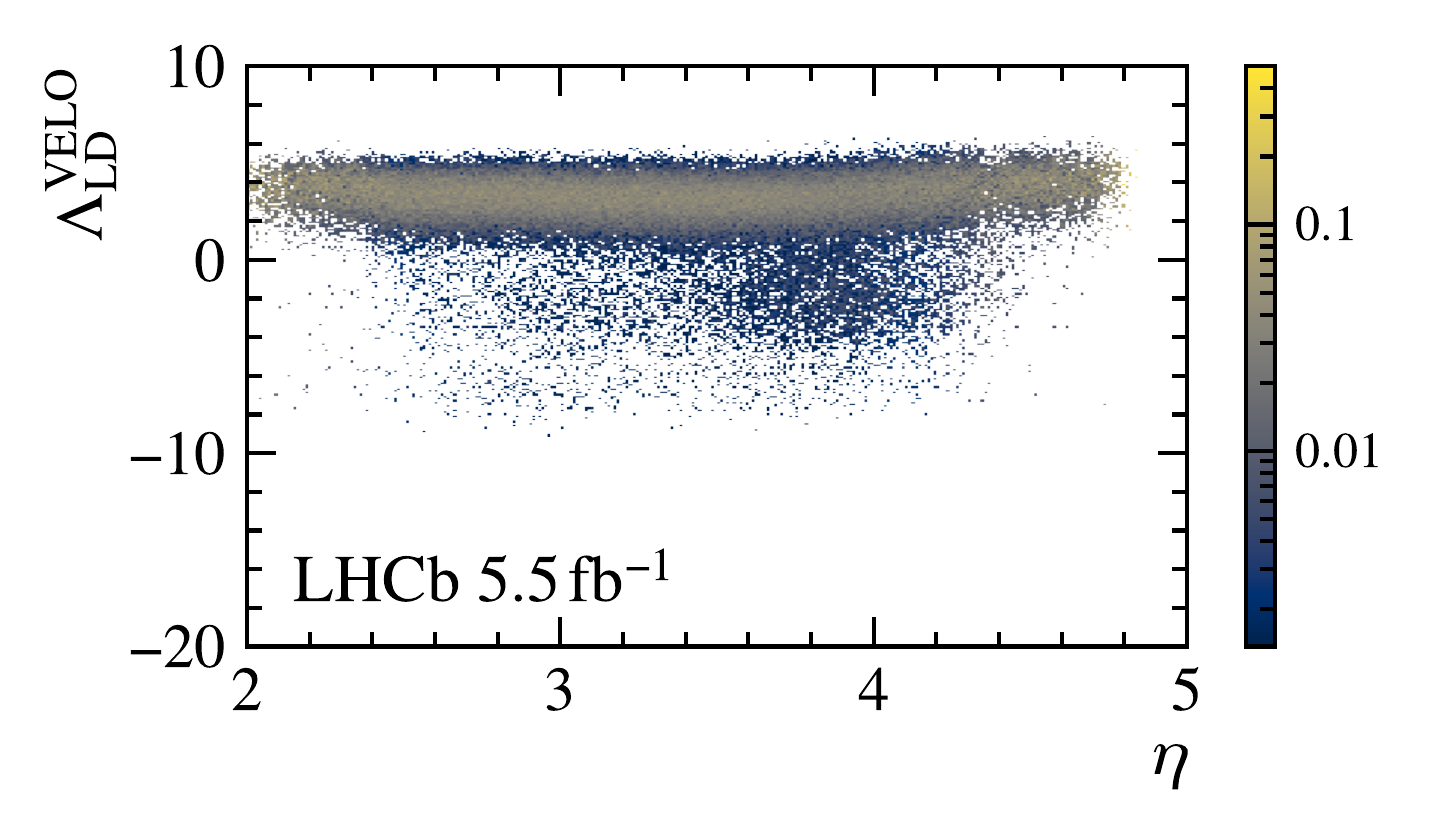} 
\caption{Distribution of \LLDvelo as a function of momentum (left) and pseudorapidity (right), in data from both preselections combined.
Tracks are required to pass the selection in~\Cref{sec:tracktime}, as well as $\LLDTT>-1$.
Each vertical slice is normalised to unity.}
\label{fig:LLDsKinematics}
\end{figure}

The \LLDvelo, \LLDTT, and \LLDIT estimators are expected to have a mass, a momentum and an incidence-angle dependence, due to the Bethe-Bloch formula. This is studied in detail, and is found to be barely resolvable with the 7-bit \adc resolution of the LHCb silicon trackers. 
As a result, the obtained distributions are continuous and nearly independent of these quantities. This is illustrated for \LLDvelo in~\Cref{fig:LLDsKinematics}.
The material in the \lhcb detector is {inhomogeneously} distributed in pseudorapidity and azimuthal angle, however no significant dependence of the separation power on these quantities is observed.

\subsection{Helium production in the detector material}
\label{sec:ipchisq}

\begin{figure}[!htbp]\centering
\includegraphics[width=0.9\textwidth]{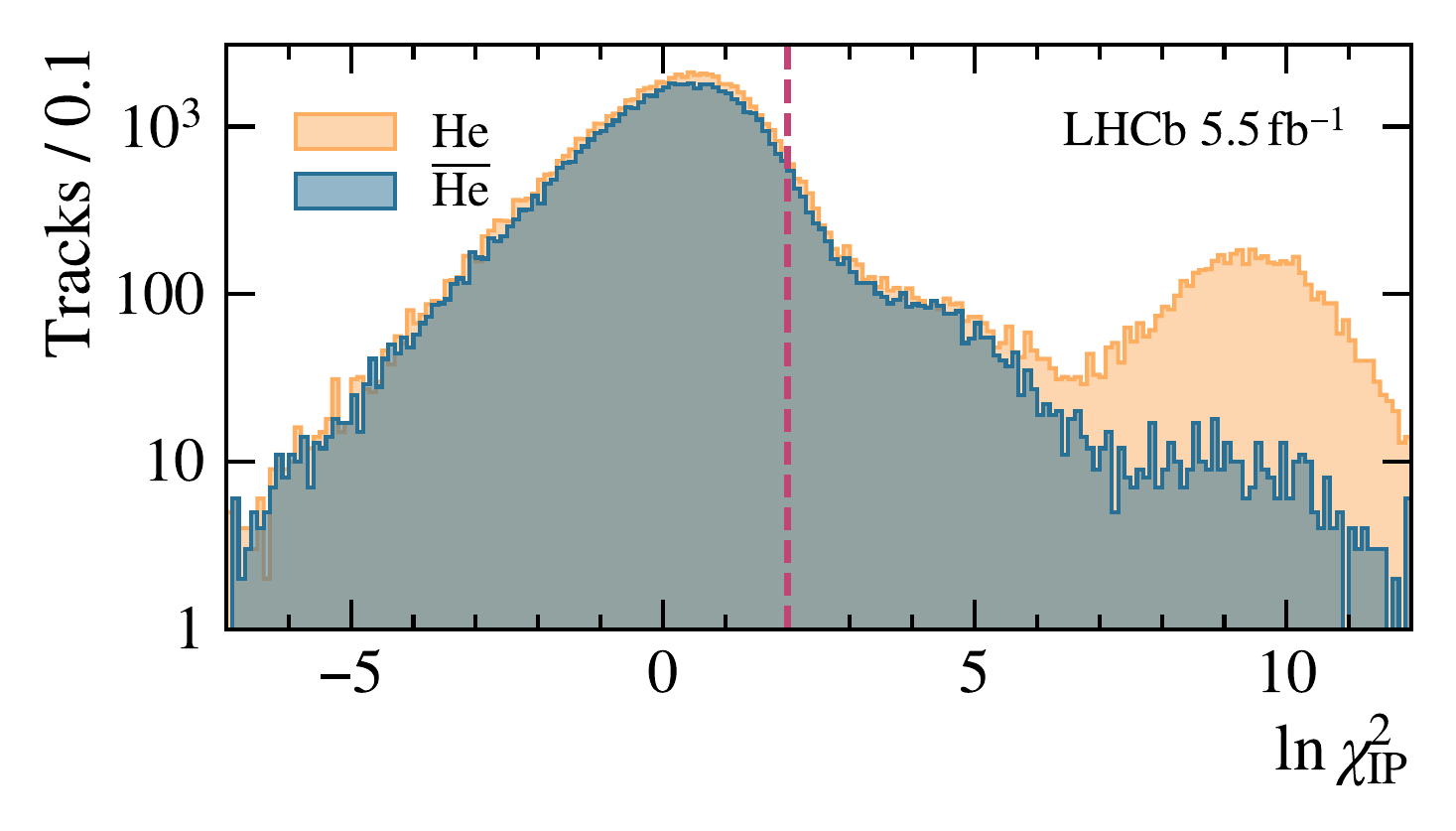} 
\caption{Distribution of the natural logarithm of the \ipchisq for helium and antihelium tracks, combining data from both preselections.
The requirement that separates prompt and displaced helium is shown {by the vertical dashed line}.}
\label{fig:heliumIPCHI2}
\end{figure}

\noindent
{Particles produced in collisions may interact with the detector material or the beam pipe, thus producing secondary particles, including helium.}
{Antihelium is not produced in this way}, thus inducing an asymmetry in distributions of variables that are different for prompt and non-prompt particles. 
One such variable is the change in the \chisq of the PV when it is reconstructed with or without the track of interest.
This quantity, denoted by \ipchisq, is very close to the square of the ratio between the impact parameter of the track and its uncertainty.

The distribution of $\ln\ipchisq$ in selected helium tracks is shown in~\Cref{fig:heliumIPCHI2}. 
An enhancement is observed at large values in the helium candidate sample, but not in the antihelium one. 
This enhancement is consistent with the production of helium in the detector material.
Therefore, prompt and displaced helium tracks are isolated by a cut on $\ipchisq$.
The upper tail in the antihelium distribution above $\ln\ipchisq>2$ is a sign of non-prompt contributions to the sample, as expected from hypertriton or \Lb decays.
Only tracks with $\ln\ipchisq<2$ are considered for the results presented throughout~\Cref{sec:results}. 

\subsection{Estimation of residual contamination}
\label{sec:bkgSubtraction}

\begin{figure}[!htbp]\centering
\includegraphics[width=0.49\textwidth]{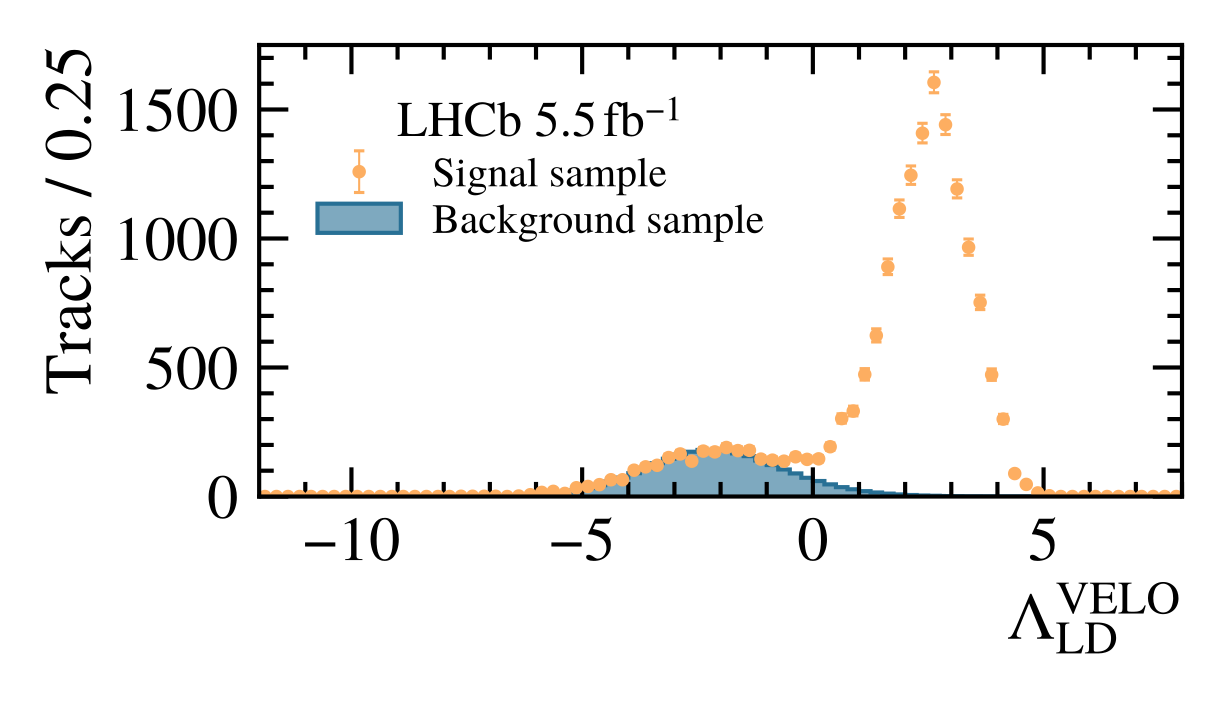} 
\includegraphics[width=0.49\textwidth]{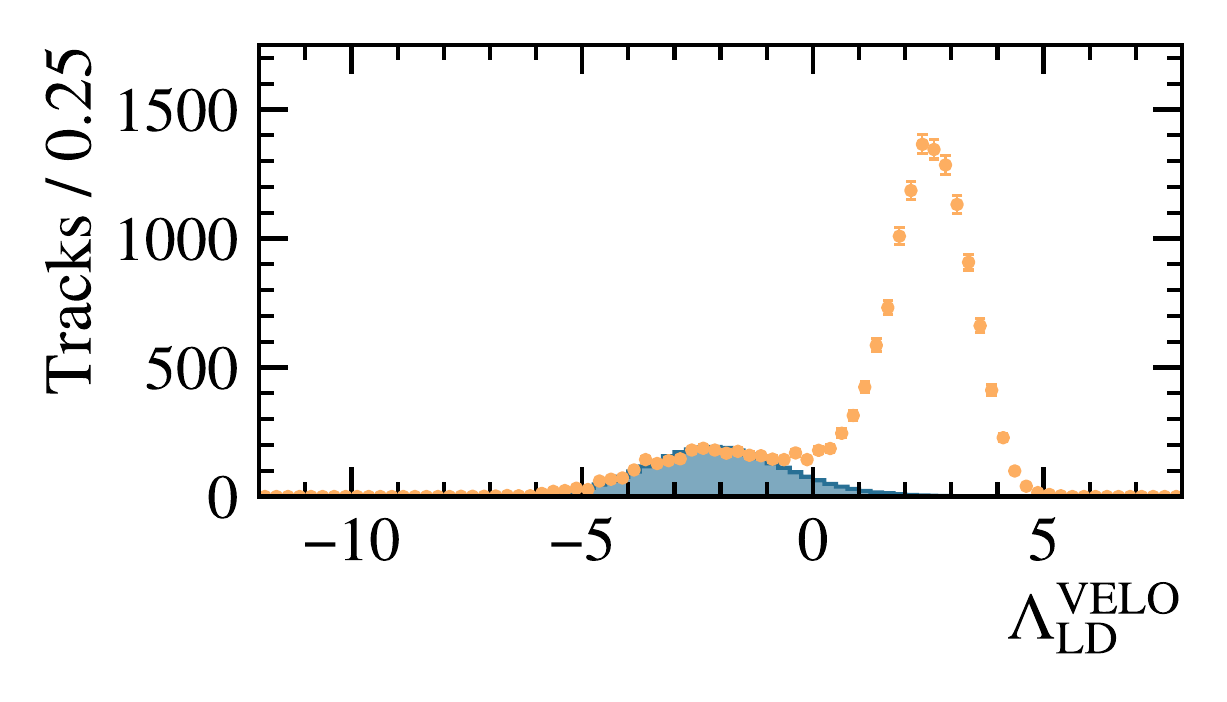} 
\includegraphics[width=0.49\textwidth]{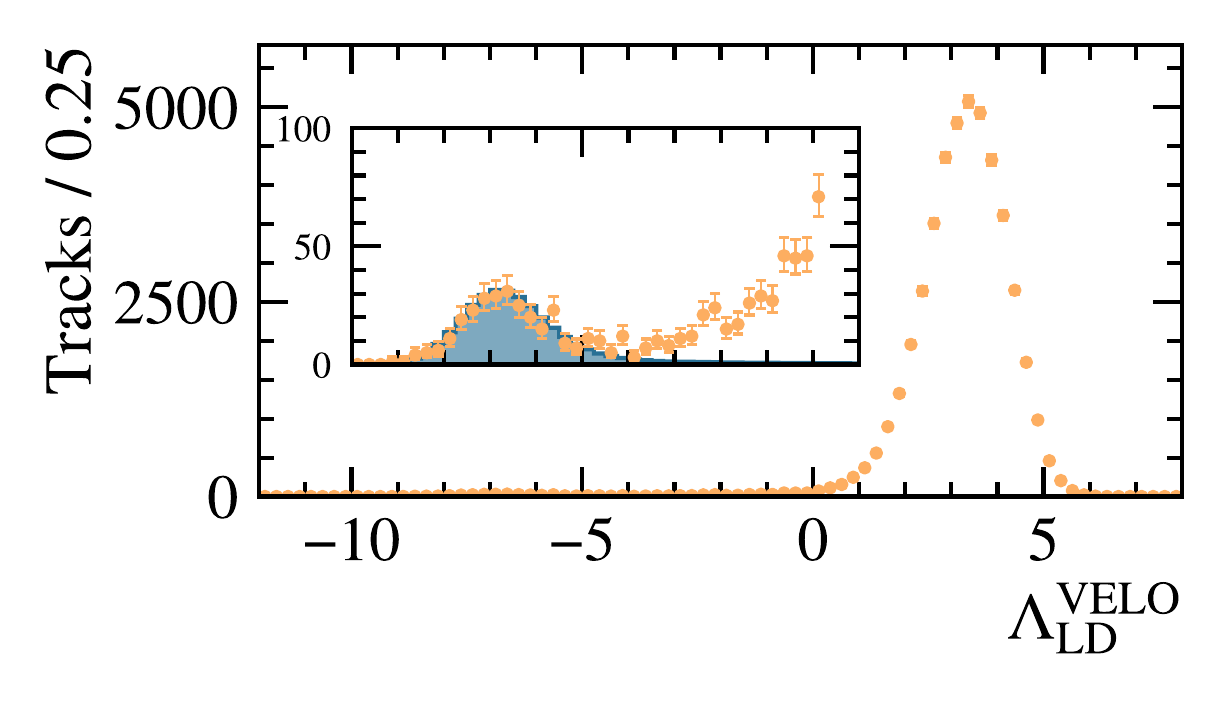} 
\includegraphics[width=0.49\textwidth]{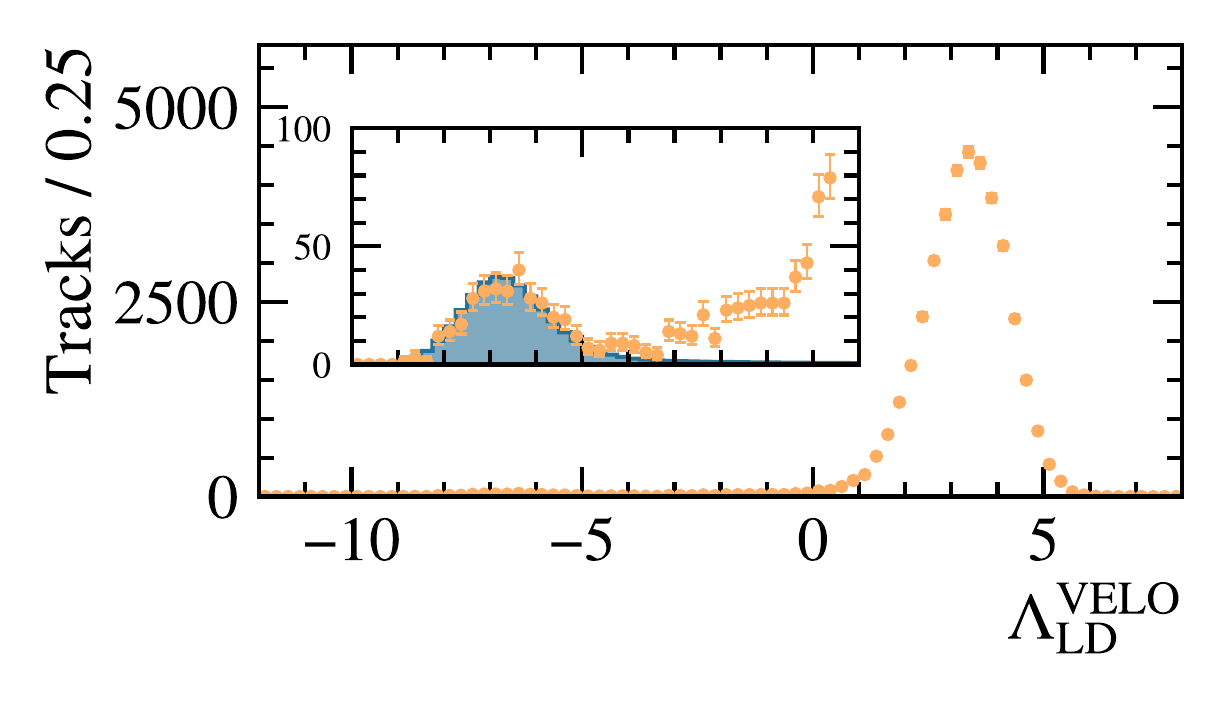} 
\caption{Distribution of \LLDvelo in (left) helium and (right) antihelium samples from the (upper) preselection~2 and (lower) preselection~1 data samples.
The distribution from an independent background sample is shown in blue, scaled to match the size of the populations found in data at negative \LLDvelo values.
To visualise the background in the preselection~1 sample, the inset of each bottom-row plot shows a magnification of the region $\LLDvelo\in[-10,\,1]$.
Each inset's horizontal axis matches that of the containing plot.}
\label{fig:heliumBinnedSigBkg}
\end{figure}

\noindent
The projections onto the \LLDvelo axis of the 2D distributions displayed in~\Cref{fig:LLDsABCD2} are used to estimate the level of residual background and are shown in~\Cref{fig:heliumBinnedSigBkg}.
The distribution of \LLDvelo in background samples obtained from regions~B~and~D ($\LLDvelo<0$) are scaled to match that in region~C.
The level of contamination is inferred to be the size of the tail at $\LLDvelo>0$.

The procedure is performed separately on the preselection~1 subset, which contains around 75\% of the tracks in region~A,
and on data from the preselection~2 sample.
This is due to the fact that the requirements for preselection~2 bias the background towards larger values, as explained {in~\Cref{sec:samples}}.
The background sample is required to have $\LLDTT<-2$.
In the case of preselection~2, the shape of the background is shifted by $+0.5$ to account for correlations between \LLDTT and \LLDvelo, which can be seen in~\Cref{fig:LLDsABCD}.
This shift is determined by minimising the $\chi^2$ between the signal and background histograms at negative \LLDvelo values.
\Cref{fig:heliumBinnedSigBkg} shows that the background in the signal region, \ie at $\LLDvelo>0$, is negligible.

The signal region of preselection~1 contains $8.7\times10^4$ tracks, of which 15 are expected to be background.
Together with the observation of 54 helium tracks in \minbias, the number of background 
tracks in the signal region in \minbias that pass the 
same requirements can be estimated to be $6\times10^{-3}$.
Given the total number of $1.2\times10^9$ well-reconstructed tracks
in the minimum-bias sample, this leads to a mis-identification probability for such a background track to pass the helium identification of $\mathcal O(10^{-12})$.

\section{Summary}
Employing techniques that are almost entirely data driven, helium and antihelium are observed for the first time at the \lhcb experiment.
This is {accomplished} using \dedx measurements in the silicon sensors (\velo, \ttracker and \intr), alongside information from the \rich and \ot subdetectors. 

A total of $1.1\times10^5$ prompt helium and antihelium are identified with negligible background contamination in the LHCb \pp collision data collected in the years 2016 to 2018.
The efficiency of the helium identification method is estimated to be approximately~$50\%$ with a corresponding background rejection factor of up to $\mathcal O(10^{12})$.

The identification method will be applied to other \lhcb \runtwo datasets, such as proton-ion, ion-ion, and SMOG collision data.
Compared to the \alice experiment, which covers the central rapidity region $|y|<0.5$,
the LHCb results will extend the available measurements in the so far experimentally unexplored forward region ($2<\eta<5$).
This identification technique, innovative at the LHCb experiment, proves the feasibility of a rich programme of measurements of QCD and astrophysics interest involving light nuclei.

\section*{Acknowledgements}
%
%
\noindent We express our gratitude to our colleagues in the CERN
accelerator departments for the excellent performance of the LHC. We
thank the technical and administrative staff at the LHCb
institutes.
We acknowledge support from CERN and from the national agencies:
CAPES, CNPq, FAPERJ and FINEP (Brazil); 
MOST and NSFC (China); 
CNRS/IN2P3 (France); 
BMBF, DFG and MPG (Germany); 
INFN (Italy); 
NWO (Netherlands); 
MNiSW and NCN (Poland); 
MCID/IFA (Romania); 
MICINN (Spain); 
SNSF and SER (Switzerland); 
NASU (Ukraine); 
STFC (United Kingdom); 
DOE NP and NSF (USA).
We acknowledge the computing resources that are provided by CERN, IN2P3
(France), KIT and DESY (Germany), INFN (Italy), SURF (Netherlands),
PIC (Spain), GridPP (United Kingdom), 
CSCS (Switzerland), IFIN-HH (Romania), CBPF (Brazil),
Polish WLCG  (Poland) and NERSC (USA).
We are indebted to the communities behind the multiple open-source
software packages on which we depend.
Individual groups or members have received support from
ARC and ARDC (Australia);
Minciencias (Colombia);
AvH Foundation (Germany);
EPLANET, Marie Sk\l{}odowska-Curie Actions, ERC and NextGenerationEU (European Union);
A*MIDEX, ANR, IPhU and Labex P2IO, and R\'{e}gion Auvergne-Rh\^{o}ne-Alpes (France);
Key Research Program of Frontier Sciences of CAS, CAS PIFI, CAS CCEPP, 
Fundamental Research Funds for the Central Universities, 
and Sci. \& Tech. Program of Guangzhou (China);
GVA, XuntaGal, GENCAT, Inditex, InTalent and Prog.~Atracci\'on Talento, CM (Spain);
SRC (Sweden);
the Leverhulme Trust, the Royal Society
 and UKRI (United Kingdom).


\addcontentsline{toc}{section}{References}
\setboolean{inbibliography}{true}
\bibliographystyle{LHCb}
\bibliography{main,standard,LHCb-PAPER,LHCb-CONF,LHCb-DP,LHCb-TDR}

\ifx\mcitethebibliography\mciteundefinedmacro
\PackageError{LHCb.bst}{mciteplus.sty has not been loaded}
{This bibstyle requires the use of the mciteplus package.}\fi
\providecommand{\href}[2]{#2}
\begin{mcitethebibliography}{10}
\mciteSetBstSublistMode{n}
\mciteSetBstMaxWidthForm{subitem}{\alph{mcitesubitemcount})}
\mciteSetBstSublistLabelBeginEnd{\mcitemaxwidthsubitemform\space}
{\relax}{\relax}

\bibitem{Winkler:2020ltd}
M.~W. Winkler and T.~Linden, \ifthenelse{\boolean{articletitles}}{\emph{{Dark matter annihilation can produce a detectable antihelium flux through $\bar{\Lambda}_b$ decays}}, }{}\href{https://doi.org/10.1103/PhysRevLett.126.101101}{Phys.\ Rev.\ Lett.\  \textbf{126} (2021) 101101}, \href{http://arxiv.org/abs/2006.16251}{{\normalfont\ttfamily arXiv:2006.16251}}\relax
\mciteBstWouldAddEndPuncttrue
\mciteSetBstMidEndSepPunct{\mcitedefaultmidpunct}
{\mcitedefaultendpunct}{\mcitedefaultseppunct}\relax
\EndOfBibitem
\bibitem{PhysRevD.103.083016}
S.~Dupourqu\'e, L.~Tibaldo, and P.~von Ballmoos, \ifthenelse{\boolean{articletitles}}{\emph{{Constraints on the antistar fraction in the Solar System neighborhood from the 10-year Fermi Large Area Telescope gamma-ray source catalog}}, }{}\href{https://doi.org/10.1103/PhysRevD.103.083016}{Phys.\ Rev.\  \textbf{D103} (2021) 083016}\relax
\mciteBstWouldAddEndPuncttrue
\mciteSetBstMidEndSepPunct{\mcitedefaultmidpunct}
{\mcitedefaultendpunct}{\mcitedefaultseppunct}\relax
\EndOfBibitem
\bibitem{AMS:2021nhj}
AMS collaboration, M.~Aguilar {\em et~al.}, \ifthenelse{\boolean{articletitles}}{\emph{{The Alpha Magnetic Spectrometer (AMS) on the international space station: Part II \textemdash{} Results from the first seven years}}, }{}\href{https://doi.org/10.1016/j.physrep.2020.09.003}{Phys.\ Rept.\  \textbf{894} (2021) 1}\relax
\mciteBstWouldAddEndPuncttrue
\mciteSetBstMidEndSepPunct{\mcitedefaultmidpunct}
{\mcitedefaultendpunct}{\mcitedefaultseppunct}\relax
\EndOfBibitem
\bibitem{STingCERNseminar}
S.~Ting, \ifthenelse{\boolean{articletitles}}{\emph{{Latest results from AMS on the International Space Station}}, }{} 8 June 2023.
\newblock \href{https://indico.cern.ch/event/1275785/}{https://indico.cern.ch/event/1275785/}\relax
\mciteBstWouldAddEndPuncttrue
\mciteSetBstMidEndSepPunct{\mcitedefaultmidpunct}
{\mcitedefaultendpunct}{\mcitedefaultseppunct}\relax
\EndOfBibitem
\bibitem{AMS:2015azc}
AMS collaboration, M.~Aguilar {\em et~al.}, \ifthenelse{\boolean{articletitles}}{\emph{{Precision measurement of the helium flux in primary cosmic rays of rigidities 1.9 GV to 3 TV with the Alpha Magnetic Spectrometer on the International Space Station}}, }{}\href{https://doi.org/10.1103/PhysRevLett.115.211101}{Phys.\ Rev.\ Lett.\  \textbf{115} (2015) 211101}\relax
\mciteBstWouldAddEndPuncttrue
\mciteSetBstMidEndSepPunct{\mcitedefaultmidpunct}
{\mcitedefaultendpunct}{\mcitedefaultseppunct}\relax
\EndOfBibitem
\bibitem{ALICE:2019vlx}
ALICE collaboration, S.~Acharya {\em et~al.}, \ifthenelse{\boolean{articletitles}}{\emph{{$^3_\Lambda\mathrm{H}$ and $^3_{\bar{\Lambda}}\mathrm{\overline{H}}$ lifetime measurement in Pb-Pb collisions at $\sqrt{s_{\mathrm{NN}}} = $ 5.02 TeV via two-body decay}}, }{}\href{https://doi.org/10.1016/j.physletb.2019.134905}{Phys.\ Lett.\  \textbf{B797} (2019) 134905}, \href{http://arxiv.org/abs/1907.06906}{{\normalfont\ttfamily arXiv:1907.06906}}\relax
\mciteBstWouldAddEndPuncttrue
\mciteSetBstMidEndSepPunct{\mcitedefaultmidpunct}
{\mcitedefaultendpunct}{\mcitedefaultseppunct}\relax
\EndOfBibitem
\bibitem{STAR:2019wjm}
STAR collaboration, J.~Adam {\em et~al.}, \ifthenelse{\boolean{articletitles}}{\emph{{Measurement of the mass difference and the binding energy of the hypertriton and antihypertriton}}, }{}\href{https://doi.org/10.1038/s41567-020-0799-7}{Nature Phys.\  \textbf{16} (2020) 409}, \href{http://arxiv.org/abs/1904.10520}{{\normalfont\ttfamily arXiv:1904.10520}}\relax
\mciteBstWouldAddEndPuncttrue
\mciteSetBstMidEndSepPunct{\mcitedefaultmidpunct}
{\mcitedefaultendpunct}{\mcitedefaultseppunct}\relax
\EndOfBibitem
\bibitem{PHENIX:2004vqi}
PHENIX collaboration, S.~S. Adler {\em et~al.}, \ifthenelse{\boolean{articletitles}}{\emph{{Deuteron and antideuteron production in $\mathrm{Au}+\mathrm{Au}$ collisions at $\sqrt{{s}_{NN}}=200\text{ }\text{ }\mathrm{GeV}$}}, }{}\href{https://doi.org/10.1103/PhysRevLett.94.122302}{Phys.\ Rev.\ Lett.\  \textbf{94} (2005) 122302}, \href{http://arxiv.org/abs/nucl-ex/0406004}{{\normalfont\ttfamily arXiv:nucl-ex/0406004}}\relax
\mciteBstWouldAddEndPuncttrue
\mciteSetBstMidEndSepPunct{\mcitedefaultmidpunct}
{\mcitedefaultendpunct}{\mcitedefaultseppunct}\relax
\EndOfBibitem
\bibitem{BRAHMS:2010nkj}
BRAHMS collaboration, I.~Arsene {\em et~al.}, \ifthenelse{\boolean{articletitles}}{\emph{{Rapidity dependence of deuteron production in central $\mathrm{Au}+\mathrm{Au}$ collisions at $\sqrt{{s}_{\mathrm{NN}}}=200$ GeV}}, }{}\href{https://doi.org/10.1103/PhysRevC.83.044906}{Phys.\ Rev.\ C \textbf{83} (2011) 044906}, \href{http://arxiv.org/abs/1005.5427}{{\normalfont\ttfamily arXiv:1005.5427}}\relax
\mciteBstWouldAddEndPuncttrue
\mciteSetBstMidEndSepPunct{\mcitedefaultmidpunct}
{\mcitedefaultendpunct}{\mcitedefaultseppunct}\relax
\EndOfBibitem
\bibitem{Kolesnikov:2015ola}
NA49 collaboration, V.~I. Kolesnikov, \ifthenelse{\boolean{articletitles}}{\emph{{Production of (anti)deuterons in heavy-ion collisions at SPS energies}}, }{}\href{https://doi.org/10.22323/1.225.0075}{PoS \textbf{Baldin ISHEPP XXII} (2015) 075}\relax
\mciteBstWouldAddEndPuncttrue
\mciteSetBstMidEndSepPunct{\mcitedefaultmidpunct}
{\mcitedefaultendpunct}{\mcitedefaultseppunct}\relax
\EndOfBibitem
\bibitem{LHCb-DP-2008-001}
LHCb collaboration, A.~A. Alves~Jr.\ {\em et~al.}, \ifthenelse{\boolean{articletitles}}{\emph{{The \lhcb detector at the LHC}}, }{}\href{https://doi.org/10.1088/1748-0221/3/08/S08005}{JINST \textbf{3} (2008) S08005}\relax
\mciteBstWouldAddEndPuncttrue
\mciteSetBstMidEndSepPunct{\mcitedefaultmidpunct}
{\mcitedefaultendpunct}{\mcitedefaultseppunct}\relax
\EndOfBibitem
\bibitem{LHCb-DP-2014-002}
LHCb collaboration, R.~Aaij {\em et~al.}, \ifthenelse{\boolean{articletitles}}{\emph{{LHCb detector performance}}, }{}\href{https://doi.org/10.1142/S0217751X15300227}{Int.\ J.\ Mod.\ Phys.\  \textbf{A30} (2015) 1530022}, \href{http://arxiv.org/abs/1412.6352}{{\normalfont\ttfamily arXiv:1412.6352}}\relax
\mciteBstWouldAddEndPuncttrue
\mciteSetBstMidEndSepPunct{\mcitedefaultmidpunct}
{\mcitedefaultendpunct}{\mcitedefaultseppunct}\relax
\EndOfBibitem
\bibitem{LHCb-TDR-005}
LHCb collaboration, \ifthenelse{\boolean{articletitles}}{\emph{{LHCb VELO (VErtex LOcator): Technical Design Report}}, }{} \href{http://cdsweb.cern.ch/search?p=CERN-LHCC-2001-011&f=reportnumber&action_search=Search&c=LHCb} {CERN-LHCC-2001-011}, 2001\relax
\mciteBstWouldAddEndPuncttrue
\mciteSetBstMidEndSepPunct{\mcitedefaultmidpunct}
{\mcitedefaultendpunct}{\mcitedefaultseppunct}\relax
\EndOfBibitem
\bibitem{LHCb-DP-2014-001}
R.~Aaij {\em et~al.}, \ifthenelse{\boolean{articletitles}}{\emph{{Performance of the LHCb Vertex Locator}}, }{}\href{https://doi.org/10.1088/1748-0221/9/09/P09007}{JINST \textbf{9} (2014) P09007}, \href{http://arxiv.org/abs/1405.7808}{{\normalfont\ttfamily arXiv:1405.7808}}\relax
\mciteBstWouldAddEndPuncttrue
\mciteSetBstMidEndSepPunct{\mcitedefaultmidpunct}
{\mcitedefaultendpunct}{\mcitedefaultseppunct}\relax
\EndOfBibitem
\bibitem{LHCb-TDR-009}
LHCb collaboration, \ifthenelse{\boolean{articletitles}}{\emph{{LHCb reoptimized detector design and performance: Technical Design Report}}, }{} \href{http://cdsweb.cern.ch/search?p=CERN-LHCC-2003-030&f=reportnumber&action_search=Search&c=LHCb} {CERN-LHCC-2003-030}, 2003\relax
\mciteBstWouldAddEndPuncttrue
\mciteSetBstMidEndSepPunct{\mcitedefaultmidpunct}
{\mcitedefaultendpunct}{\mcitedefaultseppunct}\relax
\EndOfBibitem
\bibitem{LHCb-DP-2013-003}
R.~Arink {\em et~al.}, \ifthenelse{\boolean{articletitles}}{\emph{{Performance of the LHCb Outer Tracker}}, }{}\href{https://doi.org/10.1088/1748-0221/9/01/P01002}{JINST \textbf{9} (2014) P01002}, \href{http://arxiv.org/abs/1311.3893}{{\normalfont\ttfamily arXiv:1311.3893}}\relax
\mciteBstWouldAddEndPuncttrue
\mciteSetBstMidEndSepPunct{\mcitedefaultmidpunct}
{\mcitedefaultendpunct}{\mcitedefaultseppunct}\relax
\EndOfBibitem
\bibitem{LHCb-TDR-008}
LHCb collaboration, \ifthenelse{\boolean{articletitles}}{\emph{{LHCb inner tracker: Technical Design Report}}, }{} \href{http://cdsweb.cern.ch/search?p=CERN-LHCC-2002-029&f=reportnumber&action_search=Search&c=LHCb} {CERN-LHCC-2002-029}, 2002\relax
\mciteBstWouldAddEndPuncttrue
\mciteSetBstMidEndSepPunct{\mcitedefaultmidpunct}
{\mcitedefaultendpunct}{\mcitedefaultseppunct}\relax
\EndOfBibitem
\bibitem{LHCb-TDR-006}
LHCb collaboration, \ifthenelse{\boolean{articletitles}}{\emph{{LHCb outer tracker: Technical Design Report}}, }{} \href{http://cdsweb.cern.ch/search?p=CERN-LHCC-2001-024&f=reportnumber&action_search=Search&c=LHCb} {CERN-LHCC-2001-024}, 2001\relax
\mciteBstWouldAddEndPuncttrue
\mciteSetBstMidEndSepPunct{\mcitedefaultmidpunct}
{\mcitedefaultendpunct}{\mcitedefaultseppunct}\relax
\EndOfBibitem
\bibitem{LHCb-DP-2017-001}
P.~d'Argent {\em et~al.}, \ifthenelse{\boolean{articletitles}}{\emph{{Improved performance of the LHCb Outer Tracker in LHC Run 2}}, }{}\href{https://doi.org/10.1088/1748-0221/12/11/P11016}{JINST \textbf{12} (2017) P11016}, \href{http://arxiv.org/abs/1708.00819}{{\normalfont\ttfamily arXiv:1708.00819}}\relax
\mciteBstWouldAddEndPuncttrue
\mciteSetBstMidEndSepPunct{\mcitedefaultmidpunct}
{\mcitedefaultendpunct}{\mcitedefaultseppunct}\relax
\EndOfBibitem
\bibitem{LHCb-DP-2012-003}
M.~Adinolfi {\em et~al.}, \ifthenelse{\boolean{articletitles}}{\emph{{Performance of the \lhcb RICH detector at the LHC}}, }{}\href{https://doi.org/10.1140/epjc/s10052-013-2431-9}{Eur.\ Phys.\ J.\  \textbf{C73} (2013) 2431}, \href{http://arxiv.org/abs/1211.6759}{{\normalfont\ttfamily arXiv:1211.6759}}\relax
\mciteBstWouldAddEndPuncttrue
\mciteSetBstMidEndSepPunct{\mcitedefaultmidpunct}
{\mcitedefaultendpunct}{\mcitedefaultseppunct}\relax
\EndOfBibitem
\bibitem{Calabrese:2022eju}
R.~Calabrese {\em et~al.}, \ifthenelse{\boolean{articletitles}}{\emph{{Performance of the LHCb RICH detectors during LHC~Run~2}}, }{}\href{https://doi.org/10.1088/1748-0221/17/07/P07013}{JINST \textbf{17} (2022) P07013}, \href{http://arxiv.org/abs/2205.13400}{{\normalfont\ttfamily arXiv:2205.13400}}\relax
\mciteBstWouldAddEndPuncttrue
\mciteSetBstMidEndSepPunct{\mcitedefaultmidpunct}
{\mcitedefaultendpunct}{\mcitedefaultseppunct}\relax
\EndOfBibitem
\bibitem{ALICE:2017xrp}
ALICE collaboration, S.~Acharya {\em et~al.}, \ifthenelse{\boolean{articletitles}}{\emph{{Production of deuterons, tritons, $^{3}$He nuclei and their antinuclei in pp collisions at $\mathbf{\sqrt{{\textit s}}}$ = 0.9, 2.76 and 7 TeV}}, }{}\href{https://doi.org/10.1103/PhysRevC.97.024615}{Phys.\ Rev.\  \textbf{C97} (2018) 024615}, \href{http://arxiv.org/abs/1709.08522}{{\normalfont\ttfamily arXiv:1709.08522}}\relax
\mciteBstWouldAddEndPuncttrue
\mciteSetBstMidEndSepPunct{\mcitedefaultmidpunct}
{\mcitedefaultendpunct}{\mcitedefaultseppunct}\relax
\EndOfBibitem
\bibitem{PDG2022}
Particle Data Group, R.~L. Workman {\em et~al.}, \ifthenelse{\boolean{articletitles}}{\emph{{\href{http://pdg.lbl.gov/}{Review of particle physics}}}, }{}\href{https://doi.org/10.1093/ptep/ptac097}{Prog.\ Theor.\ Exp.\ Phys.\  \textbf{2022} (2022) 083C01}\relax
\mciteBstWouldAddEndPuncttrue
\mciteSetBstMidEndSepPunct{\mcitedefaultmidpunct}
{\mcitedefaultendpunct}{\mcitedefaultseppunct}\relax
\EndOfBibitem
\bibitem{LHCb-DP-2019-001}
R.~Aaij {\em et~al.}, \ifthenelse{\boolean{articletitles}}{\emph{{Performance of the LHCb trigger and full real-time reconstruction in Run 2 of the LHC}}, }{}\href{https://doi.org/10.1088/1748-0221/14/04/P04013}{JINST \textbf{14} (2019) P04013}, \href{http://arxiv.org/abs/1812.10790}{{\normalfont\ttfamily arXiv:1812.10790}}\relax
\mciteBstWouldAddEndPuncttrue
\mciteSetBstMidEndSepPunct{\mcitedefaultmidpunct}
{\mcitedefaultendpunct}{\mcitedefaultseppunct}\relax
\EndOfBibitem
\bibitem{Aaij:2018vrk}
R.~Aaij {\em et~al.}, \ifthenelse{\boolean{articletitles}}{\emph{{Selection and processing of calibration samples to measure the particle identification performance of the LHCb experiment in Run 2}}, }{}\href{https://doi.org/10.1140/epjti/s40485-019-0050-z}{EPJ Tech.\ Instrum.\  \textbf{6} (2019) 1}, \href{http://arxiv.org/abs/1803.00824}{{\normalfont\ttfamily arXiv:1803.00824}}\relax
\mciteBstWouldAddEndPuncttrue
\mciteSetBstMidEndSepPunct{\mcitedefaultmidpunct}
{\mcitedefaultendpunct}{\mcitedefaultseppunct}\relax
\EndOfBibitem
\bibitem{Allison:2006ve}
Geant4 collaboration, J.~Allison {\em et~al.}, \ifthenelse{\boolean{articletitles}}{\emph{{Geant4 developments and applications}}, }{}\href{https://doi.org/10.1109/TNS.2006.869826}{IEEE Trans.\ Nucl.\ Sci.\  \textbf{53} (2006) 270}\relax
\mciteBstWouldAddEndPuncttrue
\mciteSetBstMidEndSepPunct{\mcitedefaultmidpunct}
{\mcitedefaultendpunct}{\mcitedefaultseppunct}\relax
\EndOfBibitem
\bibitem{Agostinelli:2002hh}
Geant4 collaboration, S.~Agostinelli {\em et~al.}, \ifthenelse{\boolean{articletitles}}{\emph{{Geant4: A simulation toolkit}}, }{}\href{https://doi.org/10.1016/S0168-9002(03)01368-8}{Nucl.\ Instrum.\ Meth.\  \textbf{A506} (2003) 250}\relax
\mciteBstWouldAddEndPuncttrue
\mciteSetBstMidEndSepPunct{\mcitedefaultmidpunct}
{\mcitedefaultendpunct}{\mcitedefaultseppunct}\relax
\EndOfBibitem
\end{mcitethebibliography}

\newpage
\centerline
{\large\bf LHCb collaboration}
\begin
{flushleft}
\small
R.~Aaij$^{35}$\lhcborcid{0000-0003-0533-1952},
A.S.W.~Abdelmotteleb$^{54}$\lhcborcid{0000-0001-7905-0542},
C.~Abellan~Beteta$^{48}$,
F.~Abudin{\'e}n$^{54}$\lhcborcid{0000-0002-6737-3528},
T.~Ackernley$^{58}$\lhcborcid{0000-0002-5951-3498},
B.~Adeva$^{44}$\lhcborcid{0000-0001-9756-3712},
M.~Adinolfi$^{52}$\lhcborcid{0000-0002-1326-1264},
P.~Adlarson$^{78}$\lhcborcid{0000-0001-6280-3851},
C.~Agapopoulou$^{46}$\lhcborcid{0000-0002-2368-0147},
C.A.~Aidala$^{79}$\lhcborcid{0000-0001-9540-4988}
% ...

{\footnotesize \it

% ...
$^{35}$Nikhef National Institute for Subatomic Physics, Amsterdam, Netherlands\\
% ...
$^{44}$Instituto Galego de F{\'\i}sica de Altas Enerx{\'\i}as (IGFAE), Universidade de Santiago de Compostela, Santiago de Compostela, Spain\\
% ...
$^{46}$European Organization for Nuclear Research (CERN), Geneva, Switzerland\\
% ...
$^{48}$Physik-Institut, Universit{\"a}t Z{\"u}rich, Z{\"u}rich, Switzerland\\
% ...
$^{52}$H.H. Wills Physics Laboratory, University of Bristol, Bristol, United Kingdom\\
% ...
$^{54}$Department of Physics, University of Warwick, Coventry, United Kingdom\\
% ...
$^{57}$School of Physics and Astronomy, University of Glasgow, Glasgow, United Kingdom\\
$^{58}$Oliver Lodge Laboratory, University of Liverpool, Liverpool, United Kingdom\\
% ...
$^{66}$Syracuse University, Syracuse, NY, United States\\
% ...
$^{78}$Department of Physics and Astronomy, Uppsala University, Uppsala, Sweden, associated to $^{57}$\\
$^{79}$University of Michigan, Ann Arbor, MI, United States, associated to $^{66}$\\
% ...
}
\end{flushleft}

\end{document}